# The role of vehicle movement in swine disease dissemination: novel method accounting for pathogen stability and vehicle cleaning effectiveness uncertainties

Jason A. Galvis, Gustavo Machado*

Department of Population Health and Pathobiology, College of Veterinary Medicine, North Carolina State University, Raleigh, NC, USA.

**\*Corresponding author:** gmachad@ncsu.edu

**Summary**

Several propagation routes drive animal disease dissemination, and among these routes, contaminated vehicles traveling between farms have been associated with indirect disease transmission. In this study, we used near-real-time vehicle movement data and vehicle cleaning efficacy to reconstruct the between-farm dissemination of the African swine fever virus (ASFV). We collected one year of Global Positioning System data of 567 vehicles transporting feed, pigs, and people to 6,363 swine production farms in two regions. In region one, without effective vehicle cleaning (0%), vehicles connected up to 2,157 farms. Individually, vehicles transporting feed connected 2,151, pigs to farms 2,089, pigs to market 1,507, undefined vehicles 1,760, and personnel three. While region two connected 437 farms. The simulation results indicated that the contact networks were reduced the most for crew transport vehicles with a 66% reduction, followed by vehicles carrying pigs to market and farms, with reductions of 43% and 26%, respectively, when 100% cleaning efficacy was achieved. Our analysis also revealed that the farms were connected by vehicles with highly stable ASFV (stability > 0.8); these edges accounted for 5% and 47% of the contacts. The results of this study showed that even when vehicle cleaning and disinfection are 100% efficacy, vehicles are still connected to numerous farms. This emphasizes the importance of better understanding transmission risks posed by vehicles to the swine industry and regulatory agencies.

# 1. Introduction

Similar to the movement of live animals known to dominate between-farm pathogen dissemination (Green et al., 2006; Galvis, Corzo and Machado, 2022), transportation of vehicle movements is of great concern as an indirect dissemination route (Galvis, Corzo and Machado, 2022; Galvis, Corzo, Prada et al., 2022). Recent studies investigated the role of vehicles as the pathway of porcine epidemic diarrhea virus (PEDV) outbreaks (Lowe et al., 2014; Boniotti et al., 2018; Garrido-Mantilla et al., 2022); African swine fever (ASF) (Li et al., 2020; D. S. Yoo et al., 2021; Adedeji et al., 2022; Cheng and Ward, 2022); and avian influenza virus (Huneau-Salaün et al., 2020; D.-S. Yoo et al., 2021). In addition, (Dee et al., 2004; Mannion et al., 2008; Greiner, 2016; Boniotti et al., 2018; Gebhardt et al., 2022) demonstrated that infectious pathogens are found on vehicle surfaces, while others estimated the contribution of vehicles in PEDV and porcine reproductive and respiratory syndrome virus (PRRSV) (Dee et al., 2002; VanderWaal et al., 2018; Galvis, Corzo and Machado, 2022). That said, the underlying mechanisms of vehicles as disease dissemination routes remain to be examined in large-scale studies (Neumann et al., 2021; Galvis, Corzo and Machado, 2022). Thus, without access to actual vehicle movement data along with pathogen stability in vehicle environments at field conditions; and the effects of vehicle cleaning and disinfection in reducing vehicle contamination, are still challenges highlighted in better understanding the indirect contribution of vehicles in disease dissemination (Bernini et al., 2019; Neumann et al., 2021; Galvis, Corzo and Machado, 2022; Gao et al., 2023).

The extraordinary complexity and the dynamics of animal and vehicle between farm movement networks present a formidable challenge for decision-makers and producers who need to implement disease control measures, often not knowing when a new load of animals will arrive and if the farm or origin has been recently infected or not, or if a feed truck is delivering feed after being at an infected farm (G.-J. Lee et al., 2019; D. S. Yoo et al., 2021; Galvis, Corzo and Machado, 2022; Galvis, Corzo, Prada et al., 2022).

Some studies in North America and Europe utilized actual animal and vehicle movement data to reconstruct the between-farm transmission dynamics of infectious diseases (Bernini et al., 2019; Andraud et al., 2022; Galvis, Corzo and Machado, 2022; Galvis, Corzo, Prada et al., 2022) while considering pathogen stability at the environment and the effects of cleaning and disinfection. Even though previous studies enhanced our understanding of indirect swine disease dissemination through vehicle movements, authors identified uncertainties about the association between i) the efficacy of vehicle cleaning and disinfection and ii) factors affecting pathogen stability over their contribution in disseminating disease from farm-to-farm (Bernini et al., 2019; Andraud et al., 2022; Galvis, Corzo and Machado, 2022; Galvis, Corzo, Prada et al., 2022). Vehicle cleaning and disinfection may not effectively eliminate infectious pathogens, especially in difficult access areas, such as behind windows or gates (Mannion et al., 2008; Boniotti et al., 2018; Li et al., 2020). Therefore, it is essential to consider that several factors modulate the impact of vehicle cleaning and disinfection effectiveness, including using different disinfectants associated or not with heat, which is directly associated with the time needed for a complete truck wash (De Lorenzi et al., 2020). Similarly, the better pathogen that survives in the environment is more likely to be disseminated among farms by vehicles (Jacobs et al., 2010; Mazur-Panasiuk and Woźniakowski, 2020). Temperature, pH, humidity, and ultraviolet (UV) radiation are associated with pathogen stability (Hijnen et al., 2006; Cutler et al., 2012; Carlson et al., 2020; Espinosa et al., 2020). For example, the high temperature reduces ASF, PRRSV, PEDV, and foot-and-mouth disease stability outside the host over time (Jacobs et al., 2010; Bøtner and Belsham, 2012; Kim et al., 2018; Mazur-Panasiuk and Woźniakowski, 2020)**.**

The scarcity of vehicle movement data and the lack of network methods capable of combining contact networks, variables associated with pathogens' stability, and uncertainty of cleaning and disinfection limit our ability to understand the contribution of vehicles in disease transmission. Here, we collected GPS data of 567 vehicles transporting feed, pigs, and people to 6,363 farms. We developed a novel vehicle contact network method that considers environmental variables and vehicle cleaning and disinfection effectiveness. Thus, our goal was to reconstruct a vehicle contact network of swine companies in the U.S. while using ASFV pathogen stability profile.

## 2. Materials and methods

*2.1 Database:* In this study, we used information from two U.S. regions. Region one with 1,974 commercial swine farms managed by six swine production companies (coded hereafter A, B, C, D, E, and F), and region two with 4,389 commercial swine farms managed by 13 swine companies (coded here as G, H, I, J, K, L, M, N, O, P, Q, R, and S). Farm data includes a unique premise identification, animal capacity stratified by age, latitude, and longitude representing the farm's centroid and associated management company. In addition, enhanced on-farm Secure Pork Supply (SPS) biosecurity plans (Center for Food Security and Public Health, 2017) were used to identify the exact farm geolocations and were available for 95.8% and 29.5% of farms located in regions one and two, respectively (subsection 2.2). Furthermore, farms were classified into 24 types based on the swine production phase or how each production company classified them. Briefly, in North American swine production, a site may have more than one production phase (i.e. farrow-to-finisher). Thus, farms are categorized based on the farm capacity of each production phase present per site. Swine companies usually have their farm classification but present inconsistencies by multiple formats among the companies. Because of this inconsistency, we simplified farm-type classification. For example, a farm with breeding-age animals was classified as a sow farm, while a farm that reported space for breeding animals and finishers was considered a sow-finisher farm (Supplementary Material Table S1 for the complete list of farm types). In regions one and two, 16% and 20% of farms, respectively, lacked pig capacity information for each production phase. For those farms as an alternative, we used farm types provided by participating companies (Supplementary Material Table S1).

Data on the vehicles used by companies A, B, and G for 2020 (from January 01 to December 31) was collected. A total of five types of vehicles were included in the study. Company A operated with 398 vehicles which included: (i) 230 trucks delivering feed to farms, named hereafter "feed-vehicle"; (ii) 169 vehicles utilized in the transportation of live pigs between farms, named hereafter "pig-farm-vehicle;" (iii) 127 vehicles used in the transportation of pigs to markets (a.k.a. slaughterhouse, packing plants) named hereafter "pig-market-vehicle"; (iv) 44 vehicles used in the transportation of crew members named hereafter

"crew-vehicle," which correspond to the movement of personnel performing a wide range of farm tasks: vaccination, power washing at closeouts, pig loading, and unloading; and (v) 84 vehicles without a defined role was named hereafter "undefined-vehicle." For company B, 105 vehicles were tracked, including 41 feed vehicles, 19 pig-farm-vehicles, 30 pig-market-vehicles, and 15 crew-vehicles. Company G 64 vehicles were monitored, and all were classified as undefined-vehicles roles. From each vehicle, 12 months of daily GPS tracker records were collected, which comprised geographic coordinates for every five seconds of any vehicle in movement. In addition, each vehicle movement included a unique identification number, speed (in km/h), date, and time. We also gathered information on 14, 3, and 15 "company-owned cleaning stations" (CCS) from companies A, B, and G, respectively. Each CCS included centroid coordinates (latitude and longitude), address, and name.

*2.2 On-farm biosecurity data:* We extracted enhanced SPS biosecurity plans data from the Rapid Access Biosecurity (RAB) application (RABapp™) database (Machado et al., 2023). Briefly, the RABapp™ serves as a platform for standardizing the approval of SPS-enhanced biosecurity plans while storing and analyzing animal and semen movement data. SPS biosecurity plans are part of a USDA and Pork Checkoff initiative (https://www.securepork.org/) to enhance business continuity by helping swine producers implement enhanced on-farm biosecurity measures on individual farms. An SPS biosecurity plan encompasses 169 unique biosecurity measures (written component) and farm maps (Center for Food Security and Public Health, 2017; Machado et al., 2023). Each farm map (Supplementary Material Figure S1) is formed of twelve biosecurity features, one of which is the Perimeter Buffer Area (PBA) is an outer control boundary around the line of separation to limit possible contamination near animal housing. It is not rare for farms to have more than one PBA because of how swine barns are distributed at a premise (Supplementary Material Figure S1). Therefore, because our methodology measures vehicle contacts to a group of barns within PBA in farms with more than one PBA, we created a unique "farm unit" identification to measure vehicle contact to each group of barns (Supplementary Material Figure S1). Our final farm population database for region one consisted of 2,519 farm units, of which 2,437 (96.7%) used PBA's geolocation, while 82 (3.3%) farms did not have an on-farm biosecurity plan, we used farm's centroid geolocation provided by the companies as an alternative. Region two consisted of 4,619 farm units with 1,523 (33%) PBA's geolocation and 3,096 (67%) farms in which we used farm centroid geolocation due to the lack of on-farm biosecurity plans.

**2.3 Vehicle movement network**

*2.3.1 Vehicle farm visit:* We defined a farm visit as a risk event in which vehicles pose a significant risk of disease introduction (Guinat et al., 2016; Li et al., 2020; Neumann et al., 2021; Galvis, Corzo and Machado, 2022). Thus, a vehicle visit was registered when a vehicle stopped within a defined distance from a "farm unit," named hereafter as "vehicle buffer distance" (VBD). In this study, we used three VBD sizes 50, 100, and 300 meters. The VBD sizes were defined based on the average length of transportation vehicles used in swine production, which ranges from 12.5 meters to 53.5 meters (Walton et al., 2009). In addition, we also tracked the time vehicles spent at VBD and conditioned a vehicle visit according to a minimum elapsed time inside that area. This time was named "vehicle visit time" (VVT) (Figure 1). It is worth noting that in some regions, third-party vehicles will deliver, for example, feed to farms of different companies. Because between-farm vehicle movements could be associated with disease dissemination among companies, we also computed the contacts among companies A, B, and G to farm units from companies C, D, E, and F in region one and H, I, J, K, L, M, N, O, P, Q, R, and S in region two.

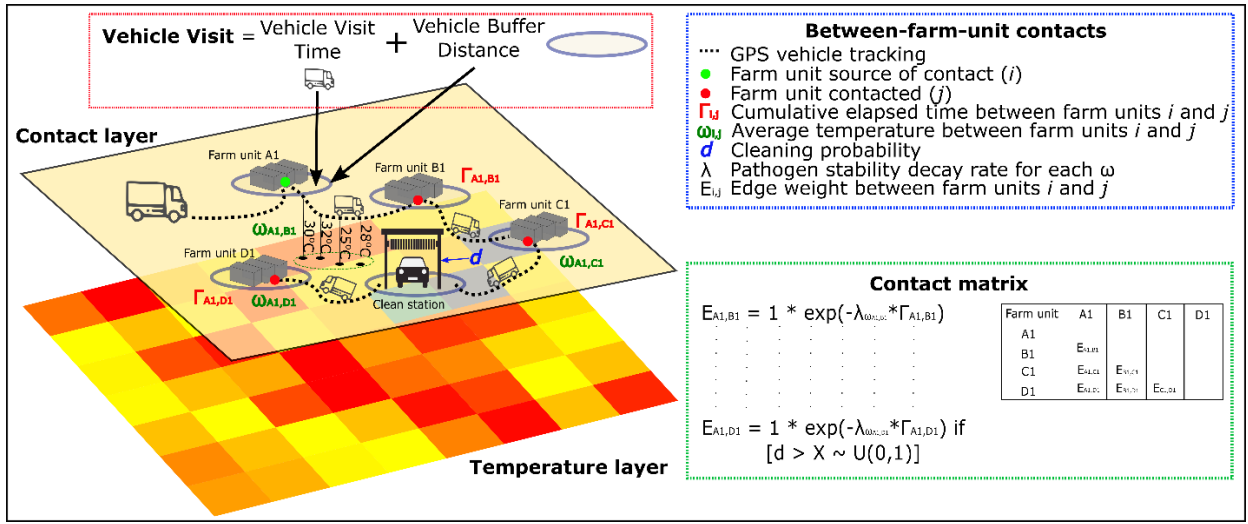

**Figure 1. Network reconstruction framework.** A vehicle visited a farm if the vehicle's latitude and longitude were inside a buffer distance and for a minimum time specified by the visit time (red box, top-left panel). An edge connecting different farms is recorded if all farms were visited by the exact vehicle in decreasing chronological order and if the edge weight (E), which represents our pathogen stability, is higher than 0 (green box, bottom-right panel). In the example, a vehicle visited four different farms, creating edges from A1→ B1 and A1→C1, and B1→ C1, while no edges were recorded from A1, B1, and C1 to D1 because the vehicle stopped at a C&D before visiting D1 and the cleaning probability $d$ was effective. The weight edges among the farms are calculated through an exponential distribution, where $\lambda$ is the decay rate for each average temperature ($\omega$) from the source of the contact (e.g., farm A1, green dot) until the destination (e.g., farms B1 and C1, red dots). Similarly, $\Gamma$ is the cumulative time from the source of the contact (e.g., farm A1, green dot) until the destination (e.g., farms B1 and C1, red dots).

*2.3.2 Farm-to-farm contact network:* We assumed a vehicle is contaminated after visiting a farm unit (Dee et al., 2002; Bernini et al., 2019), with the potential to propagate pathogens into the subsequently visited farm units. Thus, in chronological order, we computed indirect contacts among farm units visited by each vehicle and referred to these contacts as edges (*E*) (Figure 1). While we considered a range of VBD and VVT in the vehicle visit for the farm-to-farm contact network, we only evaluated the results of VBD of 50 meters and VVT of five minutes due to computational resources.

*2.3.3 Pathogen stability:* For most pathogens, the stability outside the host (a.k.a. environment) decreases as temperature increases (Espinosa et al., 2020). This phenomenon has been demonstrated for PEDV (Kim et al., 2018), PRRSV (Jacobs et al., 2010), and ASFV (Carlson et al., 2020; Mazur-Panasiuk and Woźniakowski, 2020; Nuanualsuwan et al., 2022). Here, we model pathogen stability decay as a function of time and pathogen exposure to environmental temperature. Thus, vehicle network edges are weighted by pathogen decay over time, as shown in Figure 1 (Nuanualsuwan et al., 2022). Briefly, edge weight between two farm units is modulated by two variables: i) the number of minutes a vehicle takes to go from one farm unit to another (Γ); and ii) the average environmental temperature the vehicle was exposed to along the route between these two farm units (ω) (Figure 1 and Supplementary Material Figure S2). We downloaded daily temperature raster layers with 1 km$^2$ resolution from Daily Surface Weather and Climatological Summaries (*daynet*) (Thornton, M.M. et al., 2022). Here, the GPS geolocation of each truck was matched with the respective daily temperature raster along its route between farm units (Figure 1). In addition, we assumed that pathogens' stability decay obeys an exponential distribution, a function of the environmental temperature decay rate and cumulative time that the pathogen was exposed to the environment modulated by $\lambda_\omega$ and Γ, respectively (Figure 1). The edge weight values range between 1 and 0, with one a high pathogen stability and 0 a low pathogen stability. To avoid edges with extremely low weights, we assumed weights <0.0006 were zero. Here, we evaluated edge weights frequency by grouping it into five categories: ">0.8 - 1", ">0.6 - ≤0.8", ">0.4 - ≤0.6", ">0.2 - ≤0.4," and ">0 - ≤0.2".

*2.3.4 Vehicle disinfection:* An effective farm vehicle visit to a CCS was when a vehicle came to a complete stop (0 km/h) within 500 meters of a CSS for at least 60 minutes (60 minutes was based on personal communication from the standard operating procedures for a large swine producing company) (Figure 1). We remark that eliminating 100% of organisms in vehicle surfaces via cleaning and disinfection is an optimistic assumption (Dee et al., 2004; Mannion et al., 2008; Deason et al., 2020). For example, 18% to 6% of disinfected vehicles tested positive for salmonella (Mannion et al., 2008), which could be translated to cleaning effectiveness of 82% and 94%, respectively. Similarly, in a PEDV study, 46% of disinfected vehicles were positive for PEDV via swab (Boniotti et al., 2018), cleaning effectiveness of 54%. Thus, here we simulated cleaning effectiveness ($d$), defined as a standard proportion of vehicles successfully disinfected after a CCS visit, with all possible values ranging from 0% to 100%.

*2.4 African swine fever network scenario:* We utilized the new network methodology developed in subsection 2.3 to simulate between-farm ASFV dissemination. The between-farm indirect dissemination of ASFV via contaminated vehicles has been described elsewhere (Neumann et al., 2021; Gebhardt et al., 2022), while recent studies evaluated ASFV stability under different temperatures (Supplementary Material Table S2) from which we extracted ASFV stability information used in our model. We used an exponential decay curve with different decay rates $\lambda$ for each temperature (see subsection 2.3.3 and Supplementary Material, Figure S2). The results of Mazur-Panasiuk et. al., 2020 were used for ASFV stability because it provided several stability metrics at different points in time that allowed us to reconstruct a robust decay stability curve (Supplementary Material Figure S2). Mazur-Panasiuk et. al., 2020 suggested that ASFV remains stable in soil for up to 9 days at 23 °C and 32 days at 4 °C, half-time was 0.44 days at 23 °C and 1.88 days at 4 °C, and 90% decay of 1.48 days at 23 °C and 6.26 days at 4 °C (Mazur-Panasiuk and Woźniakowski, 2020). We used a range of temperatures from 4 °C to 23 °C and assumed ASFV stability decay rate $\lambda$ was 0.001 at 4 °C and this rate increased by $4.48*10^{-05}$ for each temperature degree increase. Given that ASFV stability on temperatures lower than 4 °C and higher than 23 °C was not available, we assumed that environmental temperatures lower than 4 °C use the same decay rate as 4 °C, and temperatures higher than 23 °C use the same decay rate as 23 °C. Even though cleaning and disinfection procedures have been investigated in ASFV (De Lorenzi et al., 2020), the contributions of cleaning and disinfection procedures in eliminating the virus from vehicle surfaces are still to be fully demonstrated (Li et al., 2020; Neumann et al., 2021; Gao et al., 2023). Because of that, we decided to simulate a range of pathogen reductions ($d$) from 0, 10%, 50%, 80%, 90%, and 100%.

*2.5 Vehicle network outputs*

We evaluated nine vehicle visit scenarios, which included a factorial combination of three VBDs (50, 100, and 300 meters) and three VVTs (5, 20, and 60 minutes). We evaluated the ratio of farm unit visits, and cumulative time vehicles spent within farm units and at cleaning stations. The ratio of farm unit visits was calculated as the number of times each vehicle visited a farm divided by the number of times each vehicle visited a cleaning station. For the farm-to-farm contact network, we run ten repetitions for each cleaning effectiveness scenario. We used eight metrics to compare networks: network density, number of edges in the static and temporal networks, in-degree, out-degree, degree and betweenness centralization, and outgoing contact chains (Supplementary Material Table S3). In addition, for region one, we combined all vehicle movements and referred to this group as the combined-vehicle type. Results are presented by vehicle types and for each region.

## 3. Results

*3.1 Number of farm visits*: Table 1 shows that increasing the buffer distance around farm units leads to more vehicle visits while lengthening the minimum duration for a visit to count from 5 minutes to 20 or 60 minutes decreases the number of visits. These findings suggest a trade-off between buffer distance and visit frequency. In region one, the total number of vehicle visits varied between a minimum of 47,847 and a maximum of 301,774 visits (Supplementary Material Table S4), while the median by vehicle varied between 59 and 432 visits (Table 1). For region two, the total number of vehicle visits ranged from a minimum of 6,951 to a maximum of 15,094 (Supplementary Material Table S4), while the median by vehicle varied between 112 and 231 (Table 1).

**Table 1.** Show the median and the interquartile range (IQR) of farm units visited by each vehicle for one year.

|  | VVT |
| --- | --- |

| VBD | 5 minutes | 20 minutes | 60 minutes |
|---|---|---|---|
| **50 meters** | 364 (170-680) (R1) <br> 205 (148-278) (R2) | 326 (141-540) (R1) <br> 196 (143-274) (R2) | 59 (23,150) (R1) <br> 112 (83-147) (R2) |
| **100 meters** | 374 (175-703) (R1) <br> 215 (152-293) (R2) | 338 (147-552) (R1) <br> 207 (146-285) (R2) | 62 (25-157) (R1) <br> 118 (88-156) (R2) |
| **300 meters** | 432 (202-820) (R1) <br> 231 (158-319) (R2) | 378 (162-651) (R1) <br> 218 (147-309) (R2) | 70 (28-183) (R1) <br> 127 (97-166) (R2) |

(R1) = region 1; (R2) = region 2

The number of visits among vehicle types was found to vary significantly with variations in VBD (50, 100, and 300 meters) and VVT (5 and 20 minutes). The median number of visits by vehicle varied from 474 to 827 for feed-vehicles; 388 and 522 for pig-farm-vehicle; 277 and 360 pig-market-vehicle; 210 and 309 for undefined-vehicles; 2 and 8 for crew-vehicles, while undefined-vehicles in region two the median of visits was 205 and 231 farm units. Conversely, we observed a marked decrease in the number of visits across all vehicle types, particularly for feed-vehicles, when the minimum duration required for a visit to be considered a farm visit was extended to 60 minutes. Supplementary Material Figure S3 shows that the median number of feed-vehicle visits ranged from 22 to 29 farm units. We also demonstrated that vehicles visited farm units under the management of different companies. Company A owned vehicle visited a maximum of 19 farm units across different companies in region one, whereas, in region two, vehicles serving multiple companies visited a maximum of 12 farm units (Supplementary Material Table S5).

Regarding different farm, types visited with VBD of 50 meters and VVT of five minutes, finisher farm units were the most visited, with 33% of visits associated with feed-vehicles and less than 1% with crew-vehicles (Supplementary Material Figure S4). Pig-farm-vehicles made up 8.9% of all visits to nursery farm units, as shown in Supplementary Material Figure S4. Sow farm units were visited mainly by feed-vehicles (7.5%), followed by pig-farm-vehicles (7.2%), as shown in Supplementary Material Figure S4.

*3.2 Frequency of visits to clean stations*: In the scenario of 500 meters and at least 60 minutes within a truck wash in region one, the vehicles with the most visits to cleaning stations were pigs-market-vehicles and pigs-farm-vehicles (as shown in Table 2 and Supplementary Material Figure S13). The ratio of visits between clean stations and farm units showed that for each clean station visited, undefined-vehicles visited, on average, 4.4 (IQR 2.2-27.8) farm units. This was followed by feed-vehicles (2.9, IQR 2.9-10.6), pig-farm-vehicles (2.4, IQR 1.8-2.9), crew-vehicles (1.6, IQR 1.6-1.6) and pig-market-vehicles (1.3, IQR 1.2-1.5). Similar results were observed in region two, where undefined-vehicles visited, on average, 1.6 (IQR 1.3-2.1) farm units per clean station visit. Additional scenarios can be found in Supplementary Material Table S6-S10 and Figures S14-S15.

**Table 2.** Show the median and the interquartile range (IQR) number of clean stations visited by each vehicle for one year.

| Transportation role | Median |
| --- | --- |
| Vehicle transporting feed (R1) | 136 (21-359) |
| Vehicle transporting pigs to farms (R1) | 188 (99-238) |
| Vehicle transporting pigs to market (R1) | 206 (95-300) |
| Vehicle transporting crew (R1) | 5 (5-5) |
| Vehicle undefined (R1) | 39 (7-78) |
| Vehicle undefined (R2) | 138 (62-166) |

(R1) = region 1; (R2) = region 2

*3.3 The relationship between vehicle movement, the effectiveness of vehicle cleaning, and the stability of ASFV in the environment:* Our findings indicated slight fluctuation in network metrics across ten different cleaning and disinfection simulations in both study locations (Figure 2 and 3 and Supplementary Material Tables S11-S14). With a 100% cleaning efficacy, the maximum reduction of nodes was 14% of the crew-vehicle networks. On the other hand, the network constructed from vehicles transporting pigs to market

displayed the most significant reduction in static and temporal views, with 88% and 91% fewer edges, respectively. Furthermore, vehicles transporting pigs to market exhibited the most significant reduction of in-degree and out-degree, with 92% fewer adjacent neighbors in the network. Finally, for region two, undefined vehicles showed the most substantial decrease in the number of farm units in the outgoing contact chains, with a reduction of 76% in the total number of farm units that could be potentially exposed to indirect contact through vehicle movements.

Table 3. **Summary of network metrics.** Values represent the median of ten stochastic simulations for each $d$ evaluated; all $d$ values and IQR are available in Supplementary Material Tables S11-S12 and S14.

| Network metric | Vehicle type | | | | | | | |
| --- | --- | --- | --- | --- | --- | --- | --- | --- |
| | Combined-vehicles-R1 | | Feed-vehicles-R1 | | Pig-farm-vehicles-R1 | | Pig-market-vehicles-R1 | |
| Cleaning scenarios | Values with $d = 0\%$ | % decreased with $d = 100\%$ | Values with $d = 0\%$ | % decreased with $d = 100\%$ | Values with $d = 0\%$ | % decreased with $d = 100\%$ | Values with $d = 0\%$ | % decreased with $d = 100\%$ |
| Summary network metrics | | | | | | | | |
| Nodes | 2,159 | .05% | 2,151 | 0% | 2,103 | 0.1% | 1,618 | 1.6% |
| Edges static network | 1,232,684 | 36% | 1,018,941 | 31% | 207,232 | 83% | 139,786 | 88% |
| Density | 0.19 | 36% | 0.16 | 31% | 0.03 | 83% | 0.02 | 88% |
| In-degree | 476 | 36% | 338 | 28% | 63 | 84% | 38 | 92% |
| Out-degree | 477 | 37% | 336 | 29% | 61 | 88% | 38 | 92% |
| Edges temporal network | 5,583,703 | 57% | 3,846,333 | 52% | 841,987 | 84% | 359,796 | 91% |
| Outgoing contact chain | 2,157 | 1% | 2,159 | 1% | 2,089 | 26% | 1,507 | 43% |
| Summary of edge weight metrics | | | | | | | | |
| Edges of ASFV stability >0.8-1 | 339,490 | 6% | 198,947 | 8% | 87,603 | 1% | 20,226 | 16% |

| Edges of ASFV stability >0-<0.2 | 4,031,537 | 64% | 2,799,520 | 58% | 582,593 | 96% | 261,215 | 98% |

(R1) = region 1; (R2) = region 2

**Table 3.** Summary of network metrics. Values represent the median of 10 stochastic simulations for each *d* evaluated; all *d* values and IQR are available in Supplementary Material Tables S11-S12 and S14.

| Network metric | Vehicle type | | | | | |
|---|---|---|---|---|---|---|
| | Crew-vehicles-R1 | | Undefined-Vehicles-R1 | | Undefined-Vehicles-R2 | |
| Cleaning scenario | Values with d = 0% | % decreased with d = 100% | Values with d = 0% | % decreased with d = 100% | Values with d = 0% | % decreased with d = 100% |
| Summary network metrics | | | | | | |
| Nodes | 7 | 14% | 1,848 | 0.1% | 450 | 2% |
| Edges static network | 23 | 69% | 290,960 | 30% | 21,385 | 82% |
| Density | 0.000004 | 69% | 0.05 | 30% | 0.001 | 82% |
| In-degree | 0 | 0% | 100 | 54% | 0 | 0% |
| Out-degree | 0 | 0% | 100 | 54% | 0 | 0% |
| Edges temporal network | 24 | 71% | 535,563 | 30% | 128,483 | 89% |
| Outgoing contact chain | 3 | 66% | 1,760 | 0.2% | 437 | 76% |

| Summary edge weight metrics | | | | | | |
|---|---|---|---|---|---|---|
| Edges of ASFV stability >0.8-1 | 7 | 0% | 32,707 | 3% | 6,973 | 3% |
| Edges of ASFV stability >0-<0.2 | 13 | 100% | 388,136 | 34% | 95,888 | 97% |

7    (Continuation Table 3) (R1) = region 1; (R2) = region 2

Figure 2 and Supplementary material Table S11 indicate that the level of centralization remained relatively constant across the simulated cleaning effectiveness ($d$) for the combined and crew vehicle networks without any significant variation. On the contrary, as $d$ increased, a slight decrease in degree centralization was observed in vehicles transporting feed, while a more evident reduction was observed for all the other vehicle types (Figure 2). On the other hand, the betweenness centralization was mainly the same across simulated $ds$ for all vehicle types.

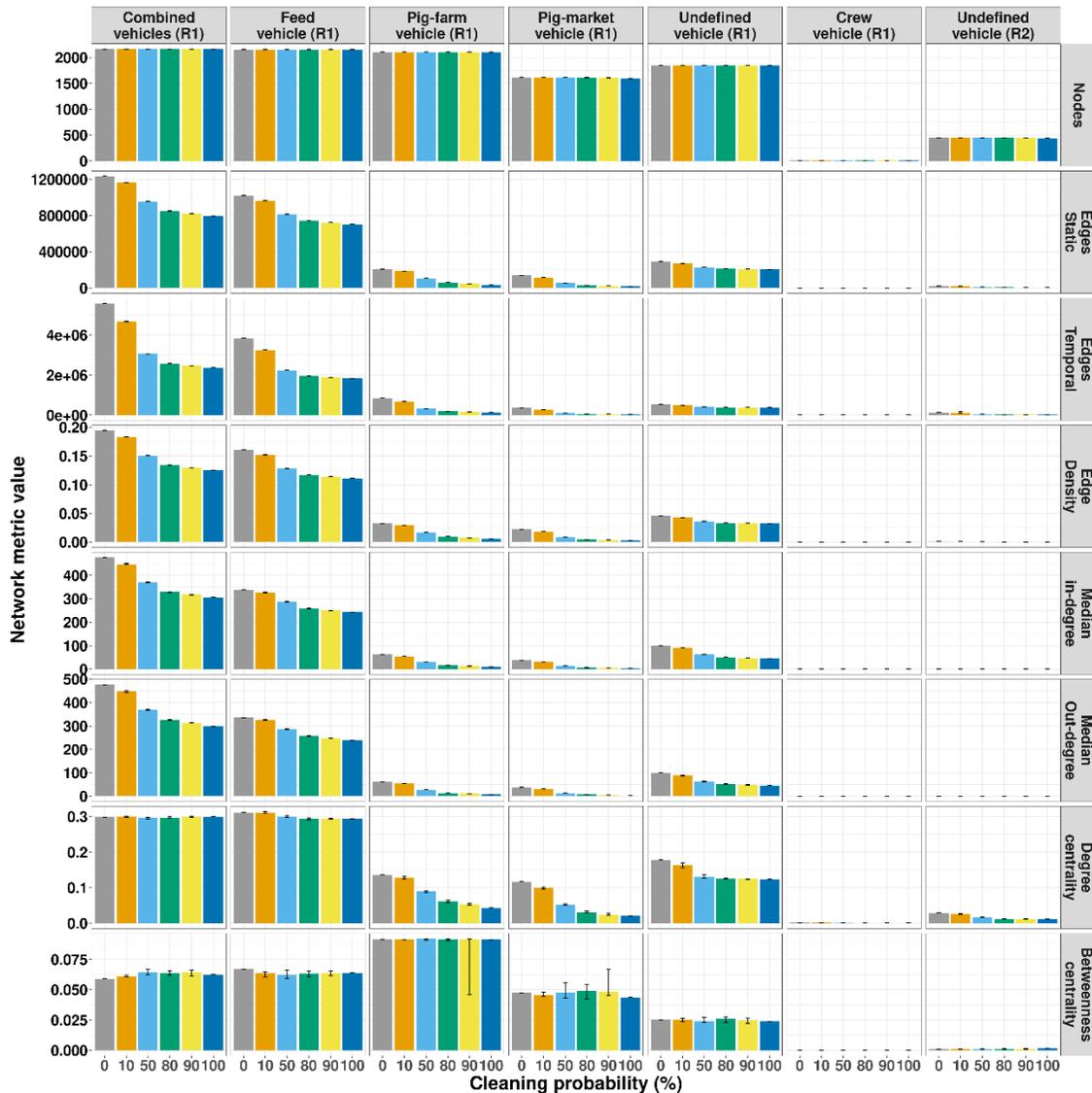

**Figure 2. Distribution of network metrics from ten different reconstructed vehicle contact networks using a VBD of 50 meters and a VVT of five minutes and six cleaning probabilities.** Bar graphs

represent the median values for each clean probability, and the error line is the minimum and maximum ranges for each distribution.

The edge weight distribution of the combined vehicles network had 6-13% of edges with ASFV stability between 0.8-1 and 61-72% of ASFV stability between 0-0.2 (Supplementary Material Table S13). In the feed vehicles network, 5% and 10% of all edges were in scenarios with ASFV stability of 0.8 to 1, while stability between 0 and 0.2, the median number of edges varied between 63% and 73%. The edges of vehicles transporting pigs with an ASFV stability >0.8 - 1 ranged from 10% to 64% and >0 - ≤0.2 from 15% to 69%. The vehicles transporting pigs to market edges ranged between 6% and 51% for ASFV stability >0.8 - 1, while the number of edges with stability between 0 and 0.2 varied between 15% and 73%. For the crew vehicle network, the median number of edges with an ASFV stability >0.8 - 1 varied between 29% and 100%, while the edges with an ASFV stability >0 - ≤0.2 varied between 0% and 54%. The undefined vehicles network in region one exhibited a median number of edges with an ASFV stability >0.8 - 1 varied between 6% and 8%, while the edges with an ASFV stability >0 - ≤0.2 varied between 68% and 72%. Finally, the network of undefined vehicles in region two showed that the median number of edges with an ASFV stability >0.8 - 1 varied between 5% and 47%, and edges with an ASFV stability >0 - ≤0.2 varied between 20% and 75%.

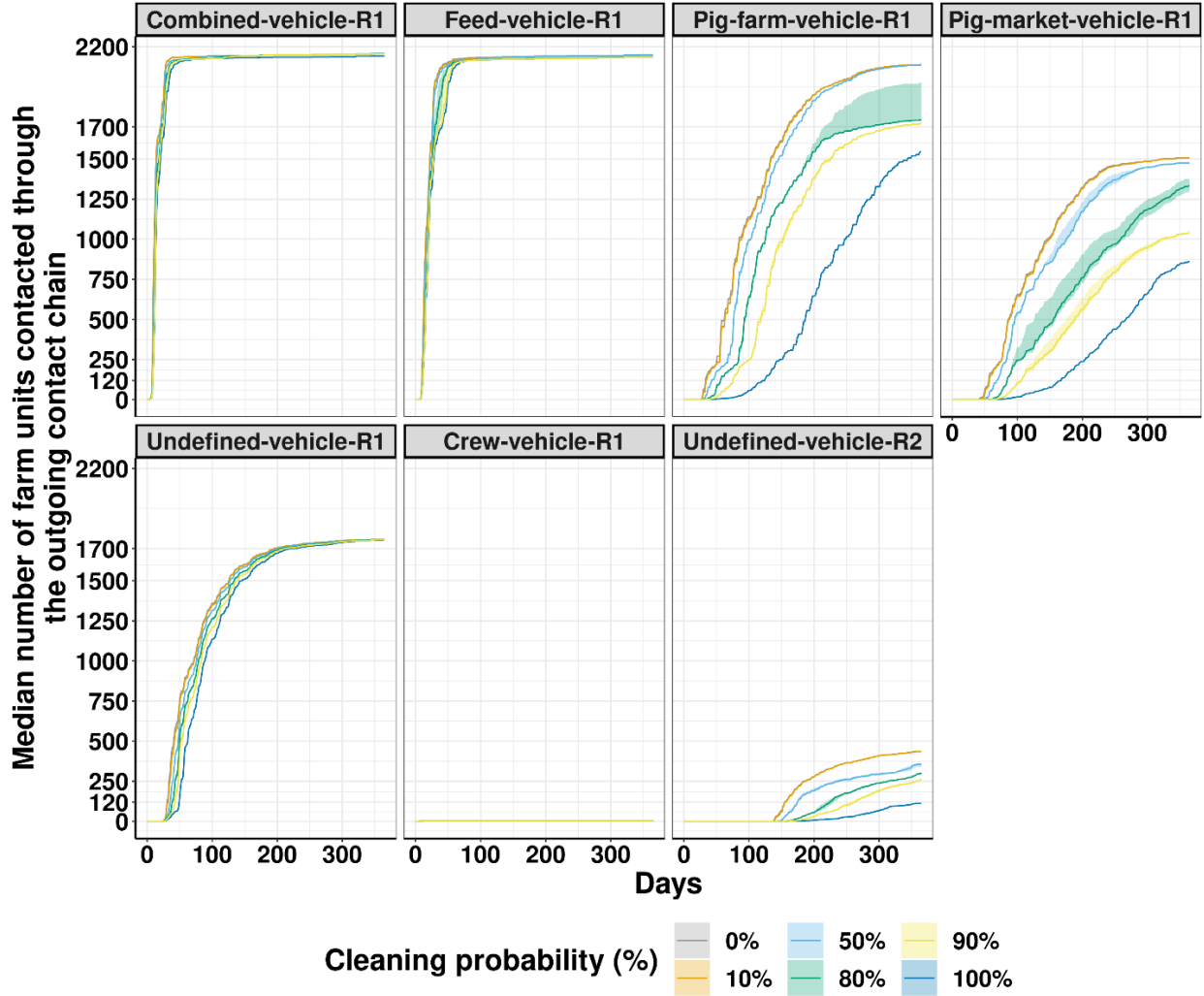

**Figure 3. The number of farm units contacted through the outgoing contact chain from vehicle movements.** Solid lines represent the median, while shadow areas represent the interquartile ranges. (R1) = region 1; (R2) = region 2.

## 4. Discussion

In this study, we developed a novel transportation vehicle contact network methodology that explicitly considers environmental pathogen stability and vehicle cleaning effectiveness uncertainties. We demonstrated that when cleaning and disinfection were either not performed in between farm visits or were simulated to be not effective ($d = 0\%$), the vehicle's contact networks had 5,583,703 edges in region one and 128,483 in region two. This means that 88% of 2,519 farm units in region one and 9% of 4,619 farm

46  units in region two were connected and potentially exposed to infected vehicles. When cleaning and
47  disinfection were simulated at 100% effective in region one, the number of edges was reduced by 57%, yet
48  87% of farm units were still connected. In region two, the number of edges was reduced by 89%, and the
49  farm units connected decreased from 9% to 2%. Additionally, for our simulated pathogen stability
50  scenarios, with $d = 100\%$, up to 13% and 47% of edges in regions one and two, respectively, were highly
51  contaminated (ASFV stability range of 0.8 to 1), thus posing significant disease spread risk. Ultimately, we
52  demonstrated that cleaning and disinfection reduced the number of edges in the vehicle to farm units'
53  movement network. Still, it was not sufficient to eliminate the contribution of vehicles spreading disease to
54  numerous farm units.

55  The frequency of visits by pig-farm-vehicles and pig-market-vehicles to cleaning stations was
56  directly related to how cleaning and disinfection disrupted their networks. This suggests that cleaning and
57  disinfection had a significant effect on disconnecting networks. We demonstrated that pig-market-vehicles
58  visit cleaning stations for every other farm visit, and pig-farm-vehicles after three farm visits. On the other
59  hand, feed-vehicles were disinfected after two and 12 farm units, and undefined-vehicles between three and
60  80 farm units (Supplementary Material Figure S15). The few cleaning feed vehicles are probably because
61  of the perceived risk of contamination from these vehicles, which do not have direct contact with animals
62  (Boniotti et al., 2018; Henry et al., 2018). Undoubtedly, pig-farm and pig-market vehicles transporting
63  animals in direct contact with infected organic material are usually recognized as high risk of disease
64  dissemination (Mannion et al., 2008; Alarcón et al., 2021). However, recent studies in Vietnam and Mexico
65  demonstrated the association between feed vehicles and ASFV dissemination (Gebhardt et al., 2022) and
66  PEDV (Garrido-Mantilla et al., 2022). Therefore, regardless of the vehicle's transportation function, we
67  emphasize the significance of increasing the frequency of disinfecting vehicles between farm visits to
68  reduce the number of indirect contacts between farms. It is essential to mention that if cleaning and
69  disinfecting are prohibited due to cost or logistic challenges (e.g., freezing weather), as described elsewhere
70  (Denver et al., 2016; Weng et al., 2016), it is recommended that efforts be made to prioritize the disinfection
71  of vehicles used for transporting animals after each farm visit at a minimum. In addition to the challenges

72    of cleaning and disinfecting, alternative strategies to minimize indirect contact between farms could involve
73    redirecting vehicles based on factors such as the health status of farms and distance, as proposed in Sweden
74    (Nöremark et al., 2009).
75    As described in recent studies (Büttner and Krieter, 2020; Galvis, Corzo and Machado, 2022),
76    transportation vehicles and animal movement network configurations differ significantly. It has beem
77    demonstrated that the vehicle transportation network links up to 100 times more farms than the animal
78    movement network (Büttner and Krieter, 2020; Galvis, Corzo and Machado, 2022), which directly impacts
79    the effectiveness of network risk-based farm ranking often proposed in disease control programs (Büttner
80    and Krieter, 2020). Our results show that vehicle transportation networks are not as pyramidal as the pig
81    movement networks, in which breeding farms on the top of the pyramid and finisher farms are at the bottom
82    (K. Lee et al., 2017; Schulz et al., 2017) (Supplementary Material Figures S17-S30). As such, because the
83    vehicle movement network configuration is more chaotic, it poses an increased risk of disease
84    dissemination, making target control strategies more challenging to be implemented (VanderWaal et al.,
85    2018; Galvis, Corzo and Machado, 2022).
86    Despite our simulated cleaning effectiveness scenarios, we uncover highly connected vehicle
87    networks, except crew vehicle networks, which connected fewer farm units. Interestingly, we observed
88    through the degree of centralization that the number of farms heavily interconnected (a.k.a. hubs) by
89    vehicles transporting pigs and undefined vehicle networks was less frequent when cleaning efficacy was
90    100%. Given that contact networks with fewer hubs have been associated with slow disease propagation
91    (Kiss et al., 2006; Martínez-López et al., 2009), increasing cleaning efficacy is indeed expected to impact
92    disease dissemination through vehicle movements.
93    Measuring the outgoing contact chain, we demonstrated that 88% of the farms were interconnected.
94    Büttner and Krieter, 2020 reported similar results in which 70% to 97% of the farms became infected via
95    transportation network. Galvis et. al., 2022, demonstrated that vehicles transporting feed significantly
96    contributed to PRRSV dissemination to breeding sites, and VanderWall et al., 2018, demonstrated that feed
97    vehicles have a high potential to introduce PEDV into new geographical areas. Although contamination and disease transmission through feed vehicles are less likely than through vehicles transporting animals, they still pose a significant risk (Büttner and Krieter, 2020; Galvis, Corzo and Machado, 2022; Galvis, Corzo, Prada et al., 2022; Sykes et al., 2022). On the contrary, pig-farm-vehicles, pig-market-vehicles, and undefined-vehicles interconnected fewer farm units. Importantly, we observed that 100% cleaning efficacy reduced by 26% and 43% the potential number of infected farm units via vehicles transporting pigs to farms and markets, respectively. It is worth mentioning that we observed significant variability in the effectiveness of reducing the number of farm units in the contact chain of pig and market vehicle networks when we simulated a cleaning efficacy of less than 100% (Figure 3). As previously discussed, these two types of vehicles are more prone to become contaminated while visiting farms. Hence, attaining a cleaning efficacy of nearly 100% is critical to mitigating the spread of diseases among swine production through contaminated vehicles that transport pigs (Mannion et al., 2008; Boniotti et al., 2018). It is worth highlighting that our findings revealed an unexpected behavior of the contact chain of the unidentified vehicle network in region two after 120 days. That was because GPS data of company G vehicles started to be collected in its entirety in May 2020 (as per personal communication).

We demonstrated that vehicles connect farm units of various swine-producing companies, thus posing a potential risk for between-company dissemination. Pathogen dissemination among swine companies is plausible and described earlier (Jara, et. al., 2020). Jara, et. al., 2020 identified distance from farms to roads as a risk for transmission, and Seedorf and Schmidt, 2017 suggested that vehicle movements may disseminate bioaerosols in the surrounding area, creating a potential infection risk for farms situated close to roads. At least in densely swine-populated regions in which the traffic of swine production related vehicles is elevated, it is likely that infectious pathogens may be circulating among swine companies, even in the absence of farm visits (Seedorf and Schmidt, 2017). Future transmission models should formally account for this novel indirect route of intercompany transmission to investigate its potential contributions to disease spread.

The outcomes of our study indicated that the majority of farm-to-farm network connections had low ASFV stability, with less than 0.2 quantity of viable virus, thereby posing a low risk of disease transmission (Carlson et al., 2020; Mazur-Panasiuk and Woźniakowski, 2020; Nuanualsuwan et al., 2022). Due to our analysis that considers the decline of ASFV stability to 90% within a maximum of 6.26 days of exposure to the environment, we expected a significant number of movements with low ASFV stability (Mazur-Panasiuk and Woźniakowski, 2020). It is noteworthy that cleaning and disinfection had a significant impact on reducing the number of contacts between farm units. When disinfection efficacy was at 100%, the edges in the combined-vehicle network with ASFV ≤0.2 reduced by 64%, whereas in the pig-market-vehicle network, this reduction was even more substantial, at 98%. Conversely, cleaning and disinfection efficacy did not substantially impact the reduction of contacts between farm units where ASFV stability exceeded 0.8. The highest reduction, with 16% fewer edges with 100% disinfection, was observed in the pig-market-vehicle network. Feed-vehicles and pig-farm-vehicles had the highest number of movements and ASFV stability greater than 0.8, even with 100% disinfection, in which pig-farm-vehicles pose an exceptionally high risk of disease dissemination due to their direct contact with animals (Alarcón et al., 2021). In conclusion, our findings suggest that enhancing the effectiveness of cleaning protocols has a minimal impact on decreasing the number of inter-farm contacts for vehicles, particularly in this simulation with elevated ASFV stability values.

## 5. Limitations and further remarks

We recognize the limitations of the novel methodology for the proposed vehicle movement network and the available vehicle movement data. It is worth noting that the absence of data from vehicles serving most, but not all, premises in both regions affected the outcomes concerning indirect contacts between companies. Likewise, we were unaware of third-party vehicle washing locations; this limitation likely impacted more significantly in crew and undefined vehicle networks, since smaller vehicles are more likely to be clean at drive-throughs at gas stations. The assumption of 60 minutes being adequate to fully clean and disinfect a vehicle may not hold in regions with freezing temperatures (Gao et al., 2023). Additionally, it should be noted that our novel network methodology utilizes GPS data from the vehicle cab and does not monitor trailers. This is because most swine companies do not track trailers via GPS; some trace trailers based on plate identification. This is a critical data limitation; however, because our methodology only requires GPS data, it can be used to reconstruct trailer networks when the data becomes available. Similarly, truck drivers with contaminated boots have been associated with disease dissemination (Dee et al., 2002). Also, Perri, et al., 2020, showed that drivers might step out of the truck at farms (Perri et al., 2020); thus, in future studies, between-farm driver movement networks should be further investigated.

As for the assumptions about the stability of pathogens, our primary limitation was that we used soil as the reference material for ASFV stability, as indicated by Mazur-Panasiuk and Woźniakowski (2020). While studies examined ASFV stability in different materials (Nuanualsuwan et al., 2022), we opted to streamline our approach by considering the data from ASFV in soil. This choice was made due to the extensive viral stability measurements investigated by Mazur-Panasiuk and Woźniakowski (2020), which enabled us to construct a more sophisticated ASFV decay curve. Similarly, we simplified the temperature effects on the ASFV stability curve due to prior evaluations of only extreme temperature ranges, including cold and warm scenarios such as 4 °C and 23 °C (Carlson et al., 2020; Mazur-Panasiuk and Woźniakowski, 2020; Nuanualsuwan et al., 2022). Despite the above mentioned limitations, this study is the first to recreate the between-farm networks using actual vehicle movement data of commercial swine companies in North America. This is also the first study that combined vehicle GPS data with pathogen environmental stability and vehicle cleaning and disinfection effectiveness. We demonstrated the potential role of vehicles in the spread of between-farm swine diseases, providing the swine industry and regulatory agencies with the necessary information to develop effective control strategies against future threats.

**6. Conclusion**

In this study, we extended a previously developed methodology for vehicle contact networks, which is commonly employed in disease transmission models (Galvis, Corzo and Machado, 2022; Galvis, Corzo, Prada et al., 2022; Sykes et al., 2022). In this updated approach, we have considered the uncertainty related to the processes of vehicle cleaning and disinfection, as well as the decay of ASFV stability in the environment. Our study revealed that although efficient cleaning and disinfection measures affected the number of farms connected through vehicle movements, simulations with 100% cleaning and disinfection still resulted in 88% of farms being in contact over one year. Importantly, achieving 100% cleaning effectiveness reduced the risk of between-farm contacts only when the ASFV stability was low (≤0.2). Conversely, there was an insignificant reduction in the number of between-farm contacts when the ASFV stability was still high (>0.8). We noted that farms of different swine production companies were visited by vehicles that also visited farms under other production companies, enhancing the potential for between-company dissemination. This study enhances our understanding of the role of transportation vehicles in spreading diseases between farms and the risks involved. The new methodology introduced in this study can be used to develop novel disease control strategies, including rerouting vehicles based on their infection status.

**Funding**

This project was funded by the Swine Health Information Center.

**CRediT authorship contribution statement**

JAG and GM conceived the study, participated in the study's design, and coordinated the data collection. JAG conducted data processing, cleaning, designing the model, and simulated scenarios. JAG designed the computational analysis. JAG and GM wrote and edited the manuscript. Both authors discussed the results and critically reviewed the manuscript. GM secured the funding.

**Declaration of Interest**

All authors confirm that there are no conflicts of interest to declare.

**Acknowledgments**

The authors would like to acknowledge participating companies.

**Data Availability**

The data that support the findings of this study are not publicly available and are protected by confidential agreements. Therefore, they are not available.

# Supplementary Material

**Section 1. Transportation vehicle movements networks and network metrics.**

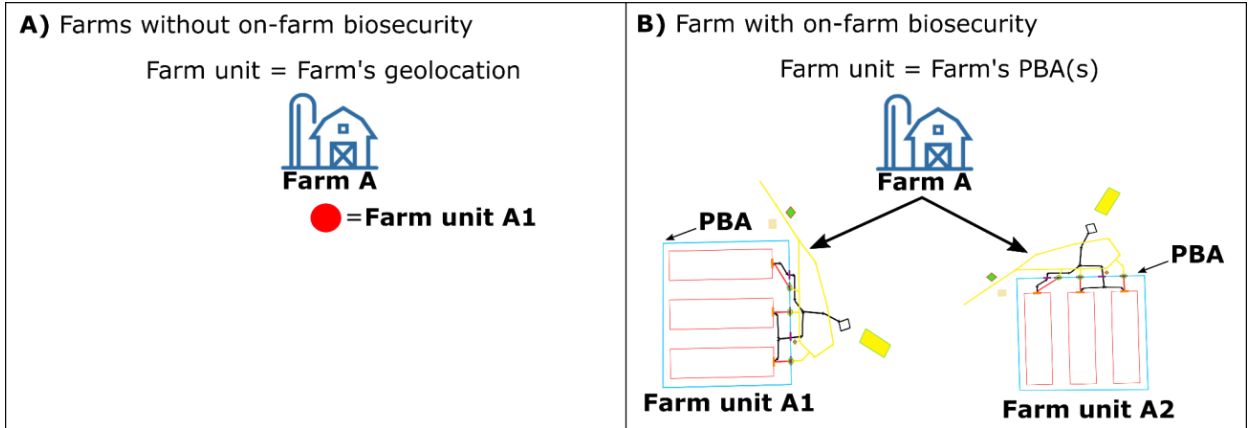

**Figure S1. Farm unit definition.** A) For farms without on-farm biosecurity plans the farm's geolocation is considered a farm unit (farm unit A1), and B) For Farms with on-farm biosecurity plans, each perimeter buffer area (PBA) geolocation (blue polygons) is considered a farm unit. Farms with multiple PBAs each receive a different farm unit identification. For example, farm A has two farm units, A1 and A2, and each one is considered as an individual node in the contact network.

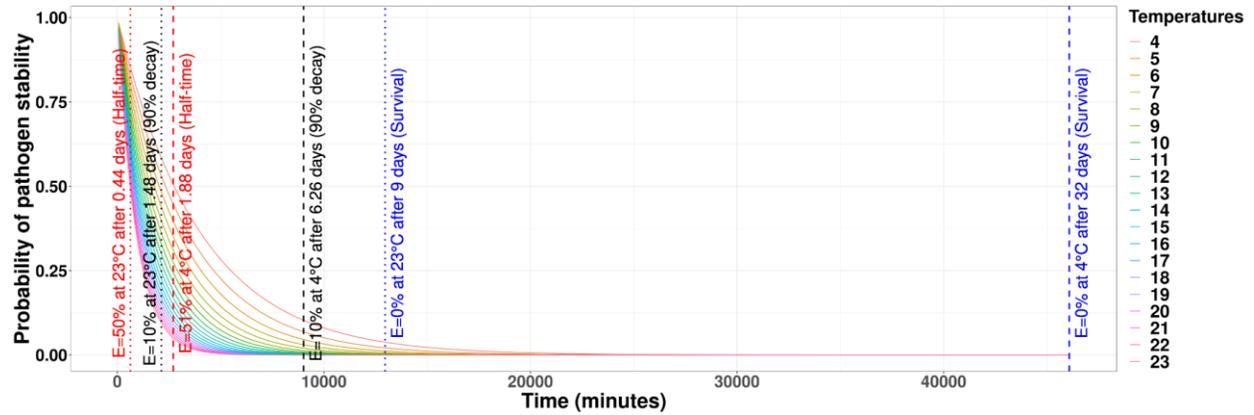

**Figure S2. Simulated relationship among temperature and time of African swine fever virus (ASFV) stability on vehicle surfaces.** Exponential decay curve for each temperature, vertical lines represent ASFV stability, half-time and D-value (90% decay) described by (Mazur-Panasiuk and Woźniakowski, 2020).

387 **Table S1.** Production types defined for the farm units in region 1 and 2.

| Farm type | Number of farms | |
| --- | --- | --- |
| | **Region one** | **Region two** |
| Boar stud | 23 | 16 |
| Finisher | 1371 | 2,265 |
| Gilt | 15 | 46 |
| Gilt-finisher | 3 | 1 |
| Gilt-nursery | 1 | 0 |
| Gilt-sow | 106 | 46 |
| Gilt-sow-nursery | 1 | 0 |
| Gilt-sow-boar stud | 0 | 35 |
| Gilt-sow-finisher-boar stud | 0 | 1 |
| Gilt-sow-nursery-finisher | 0 | 1 |
| Gilt-sow-wean to finish | 0 | 1 |
| Gilt-sow-wean to finish-boar stud | 0 | 7 |
| Isolation | 0 | 6 |
| Nursery | 546 | 403 |
| Nursery-finisher | 22 | 253 |
| Sow | 183 | 314 |
| Sow-finisher | 4 | 1 |
| Sow-nursery | 10 | 3 |
| Sow-nursery-finisher | 15 | 6 |
| Sow-nursery-finisher-isolation | 1 | 0 |
| Sow-nursery-isolation | 1 | 0 |
| Sow-wean to finish | 0 | 1 |
| Wean to finish | 214 | 1,206 |
| Wean to finish-finisher | 3 | 7 |

388
389

390 **Table S2. Stability of African swine fever virus (ASFV) at different temperatures**

| Material | Temperature (°C) | Time | Reference |
| --- | --- | --- | --- |
| Soil | 25 | 3 days | (Carlson et al., 2020) |
| | -4 | 7 days | |
| Soil | 23 | 9 days | (Mazur-Panasiuk and Woźniakowski, 2020) |
| | 4 | 32 days | |
| Soil - half-life | 23 | 0.44 days | |
| | 4 | 1.88 days | |
| Soil-decay 90% | 23 | 1.48 days | |
| | 4 | 6.26 days | |
| Metal | 25 | >7 days | (Nuanualsuwan et al., 2022) |
| | 33 | ~3 days | |
| | 42 | ~1 day | |
| Glass | 25 | >7 days | |
| | 33 | ~6 days | |
| | 42 | ~1 day | |
| Rubber | 25 | >7 days | |
| | 33 | ~5 days | |
| | 42 | ~1 day | |

391
392

393 **Table S3. Network metrics and description**

| Metric | Description | Reference |
| --- | --- | --- |
| Node | Element of the network representing the farms. | - |
| Edge | Link among two nodes. | - |
| Static network | Once an edge exists between two nodes, it is present for the whole time period. | (Kao et al., 2007) |
| Temporal network | The edges between two nodes only exist at different time steps. | (Lentz et al., 2016) |
| Density | Represent the proportion of edges among nodes in the network that are actually present. | (Wasserman and Faust, 1994) |
| In-degree | Number of nodes providing animals to a specific node. | (Wasserman and Faust, 1994) |
| Out-degree | Number of nodes obtaining animals from a specific node. | (Wasserman and Faust, 1994) |
| Degree centralization | It is a graph-level centrality score based on degree node-level centrality. It equals 1 when one node chooses all other nodes in the network (star graph). It equals 0 when all degrees are equal (circle graph) | (Wasserman and Faust, 1994) |
| Betweenness centralization | It is a graph-level centrality score based on betweenness node-level centrality. The score takes on values between 0 and 1. The maximum score is reached when one node has the largest possible betweeness, while the others the smallest possible (star graph). | (Wasserman and Faust, 1994) |
| Outgoing contact chain (OCC) | Subsets of nodes that can be reached by a specific node by direct contact or indirect contacts through a sequential order of edges through other nodes using the temporal network. | (Nöremark and Widgren, 2014) |

394

**Section 2. Vehicle visits**

**Table S4.** Number of farm units visited vehicles for one year.

|  | VVT | | |
| --- | --- | --- | --- |
| **VBD** | **5 minutes** | **20 minutes** | **60 minutes** |
| **50 meters** | 252,355 (R1)<br>13,016 (R2) | 191,349 (R1)<br>12,482 (R2) | 47,847 (R1)<br>6,951 (R2) |
| **100 meters** | 262,114 (R1)<br>13,834 (R2) | 199,012 (R1)<br>13,218 (R2) | 50,335 (R1)<br>7,365 (R2) |
| **300 meters** | 301,774 (R1)<br>15,094 (R2) | 229,747 (R1)<br>14,394 (R2) | 58,940 (R1)<br>7,944 (R2) |

(R1) = region one; (R2) = region two

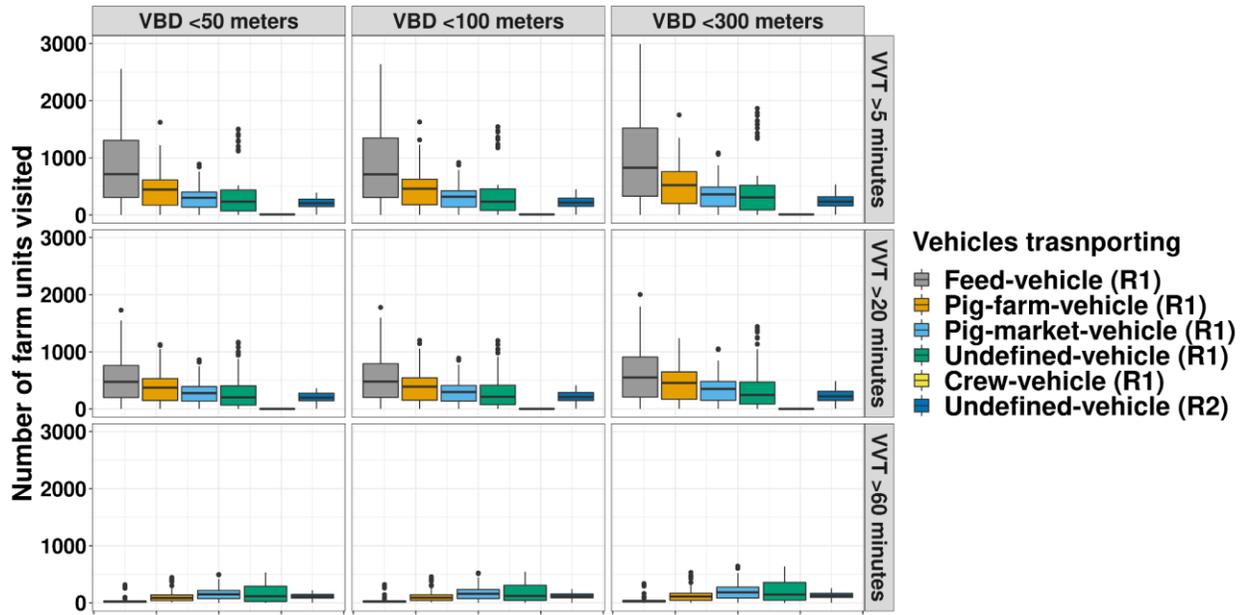

**Figure S3. Frequency of farm units visited by different vehicles from company A, B and G from January to December 2020.** The y-axis represents the number of farm units visited, box plot fill colors represent the vehicle's transportation type and are grouped by region one (R1) and region two (R2). The columns and rows represent the three different vehicle buffer distances (VBD) and three vehicle visit time (VVT).

409  **Table S5.** Number and proportion of farm units visited by vehicles from different companies

| Companies of origin | Companies of origin | VBD | VVT > 5 minutes | VVT > 20 minutes | VVT > 60 minutes |
|---|---|---|---|---|---|
| A | B-F | **<50 meters** | 11 (2.7%) | 10 (2.4%) | 6 (1.4%) |
| A | B-F | **<100 meters** | 11 (2.6%) | 10 (2.4%) | 6 (1.4%) |
| A | B-F | **<300 meters** | 16 (3.9%) | 14 (3.4%) | 7 (1.7%) |
| B | A and C-F | **<50 meters** | 2 (0.1%) | 2 (0.1%) | 2 (0.1%) |
| B | A and C-F | **<100 meters** | 3 (0.1%) | 3 (0.1%) | 2 (0.1%) |
| B | A and C-F | **<300 meters** | 19 (0.9%) | 16 (0.7%) | 13 (0.6%) |
| G | H-S | **<50 meters** | 9 (0.2%) | 9 (0.2%) | 6 (0.15%) |
| G | H-S | **<100 meters** | 10 (0.2%) | 10 (0.2%) | 7 (0.17%) |
| G | H-S | **<300 meters** | 12 (0.3%) | 11 (0.3%) | 8 (0.2%) |

410

411

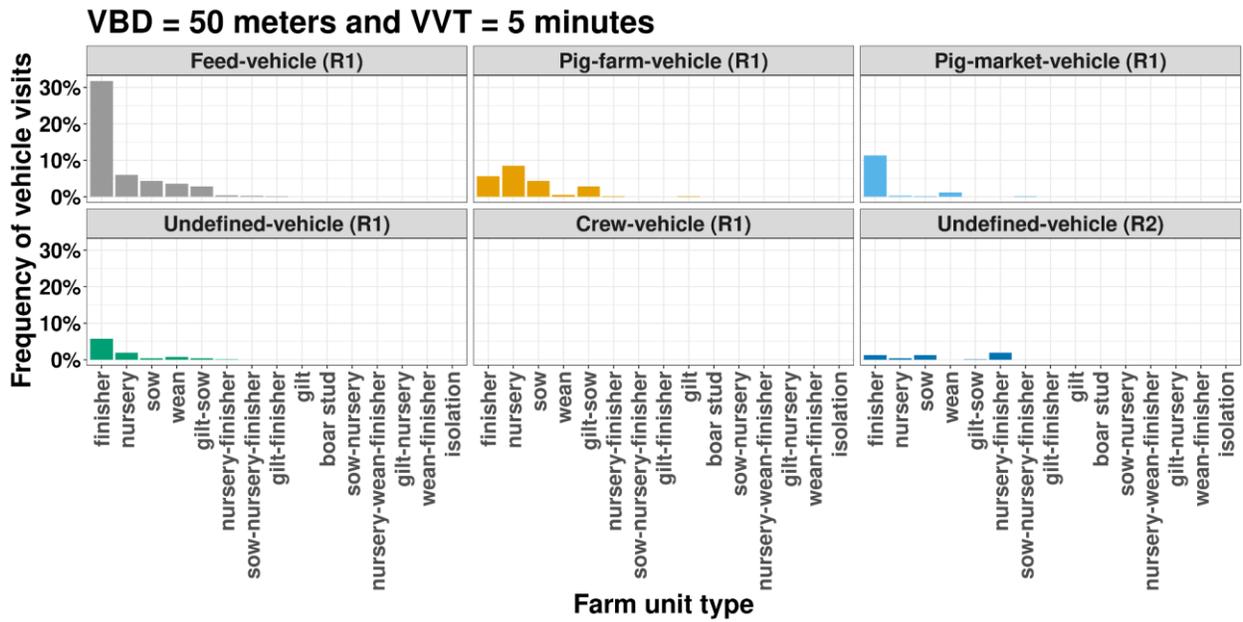

412
413 **Figure S4. Vehicle visit frequency company by farm type**
414
415

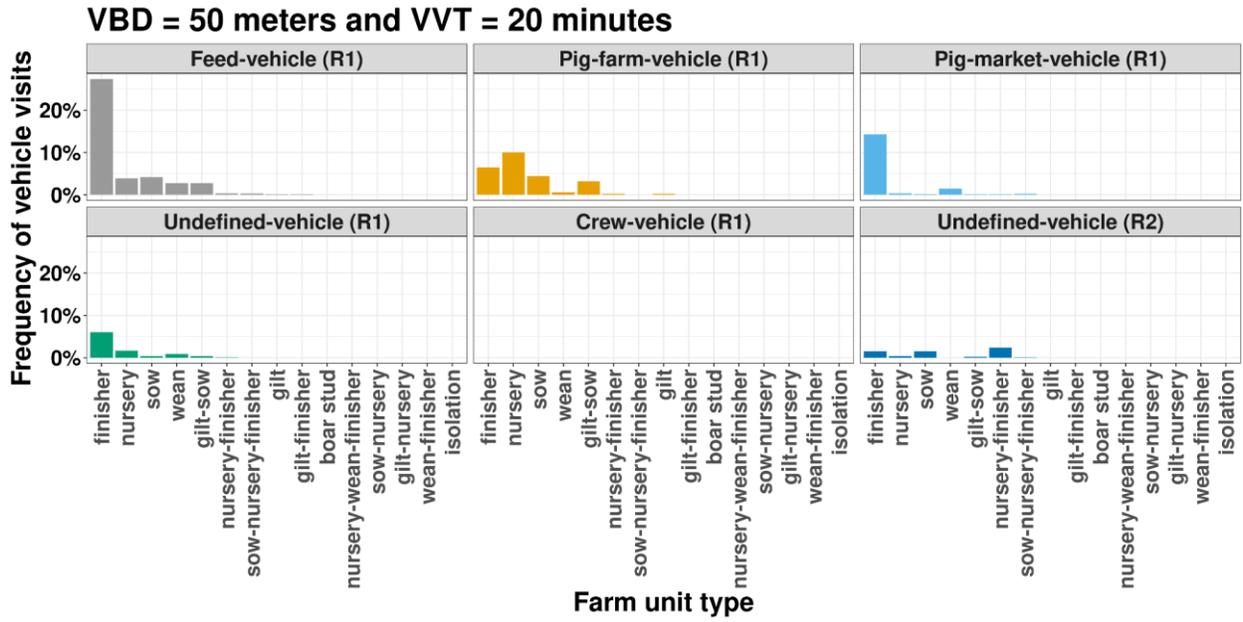

**Figure S5. Vehicle visit frequency company by farm type**

421

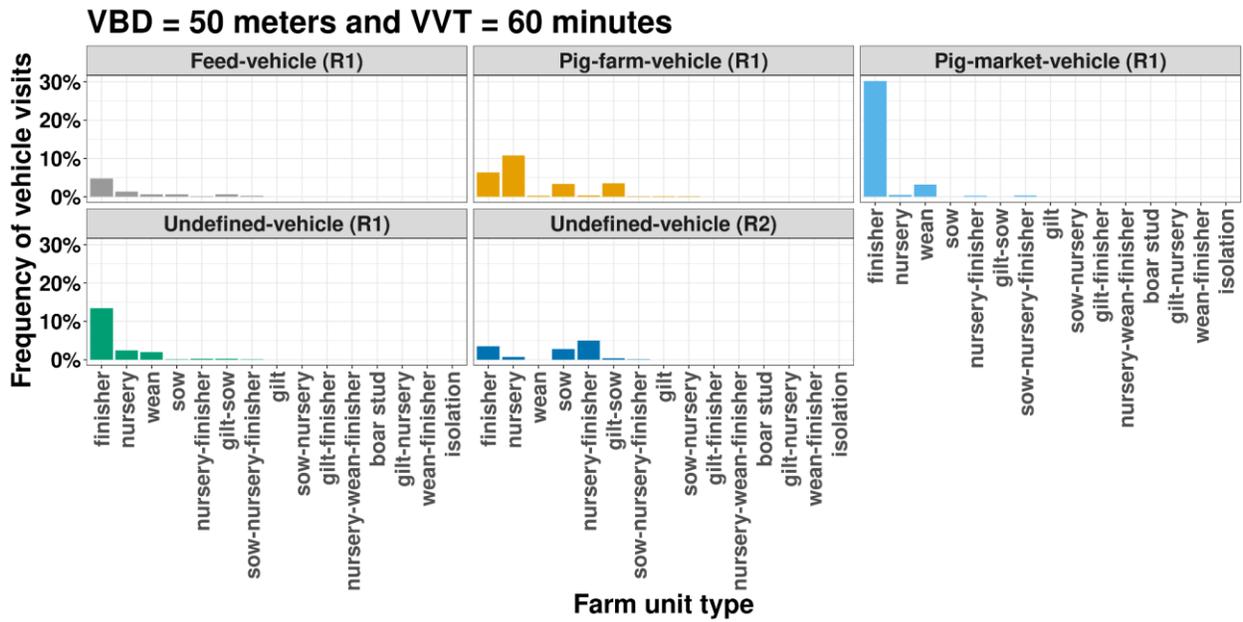

422
423 **Figure S6. Vehicle visit frequency company by farm type**
424
425

426

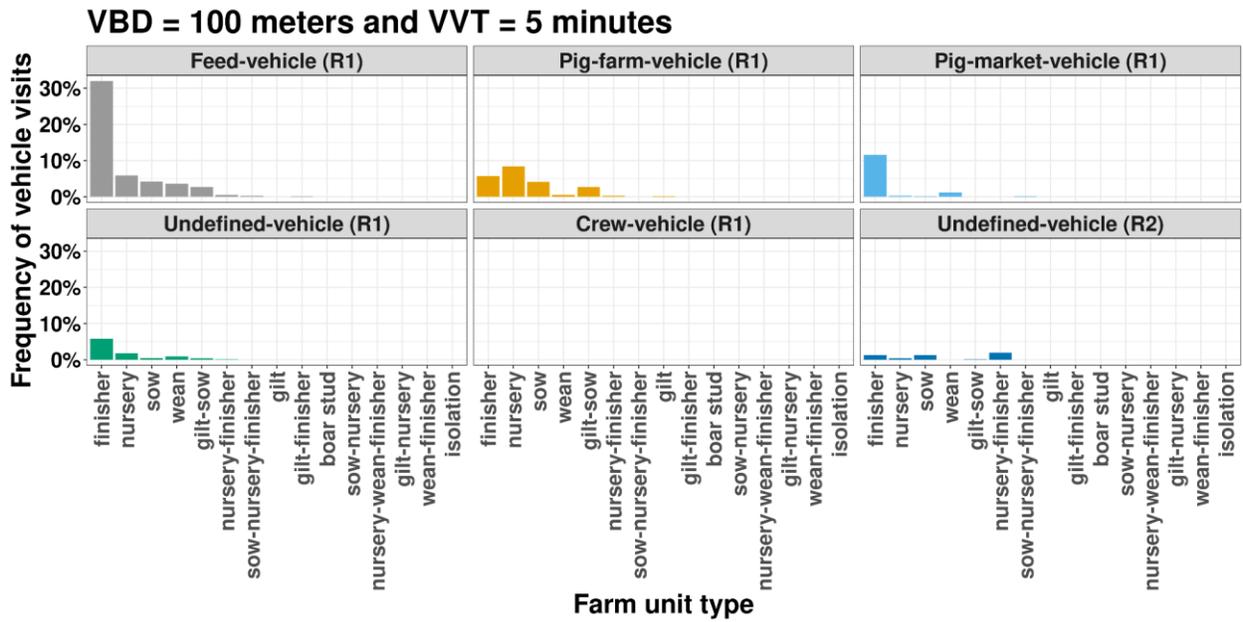

427
428 **Figure S7. Vehicle visit frequency company by farm type**
429

430

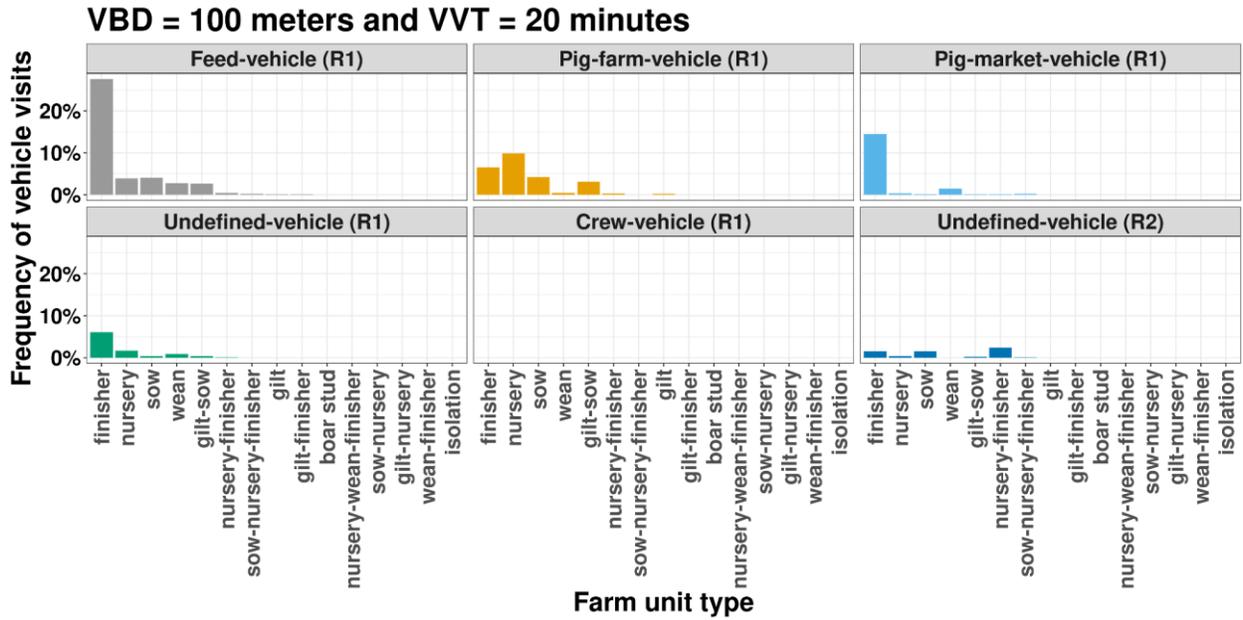

431
432 **Figure S8. Vehicle visit frequency company by farm type**
433

434

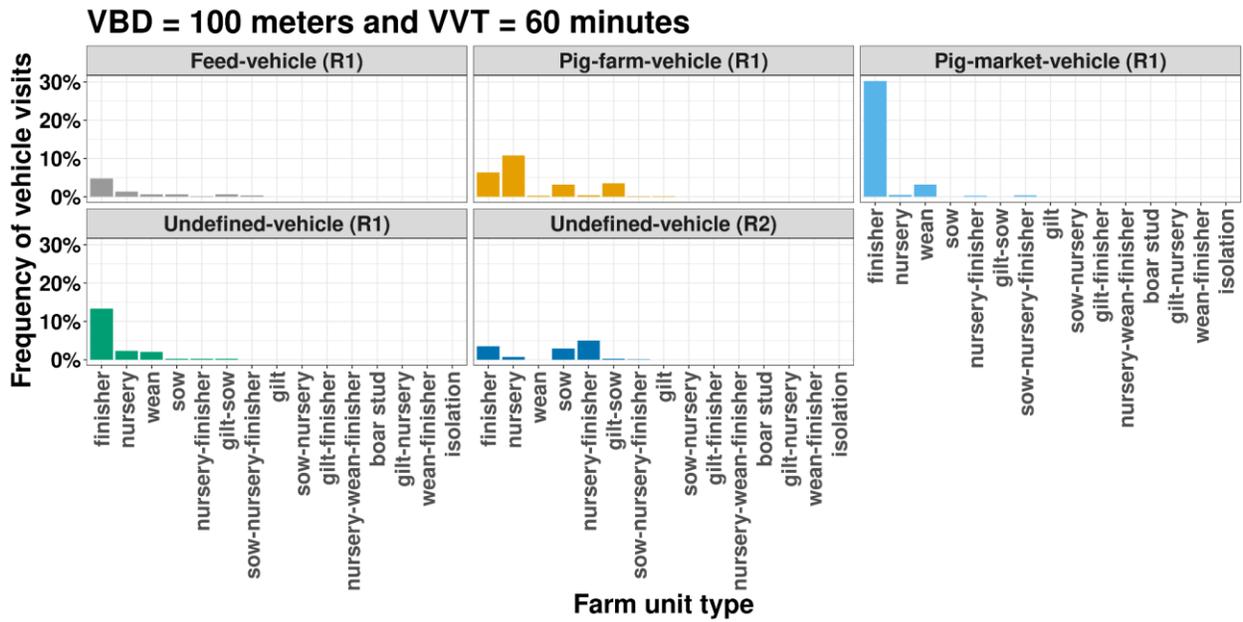

435
436 **Figure S9. Vehicle visit frequency company by farm type**
437

438

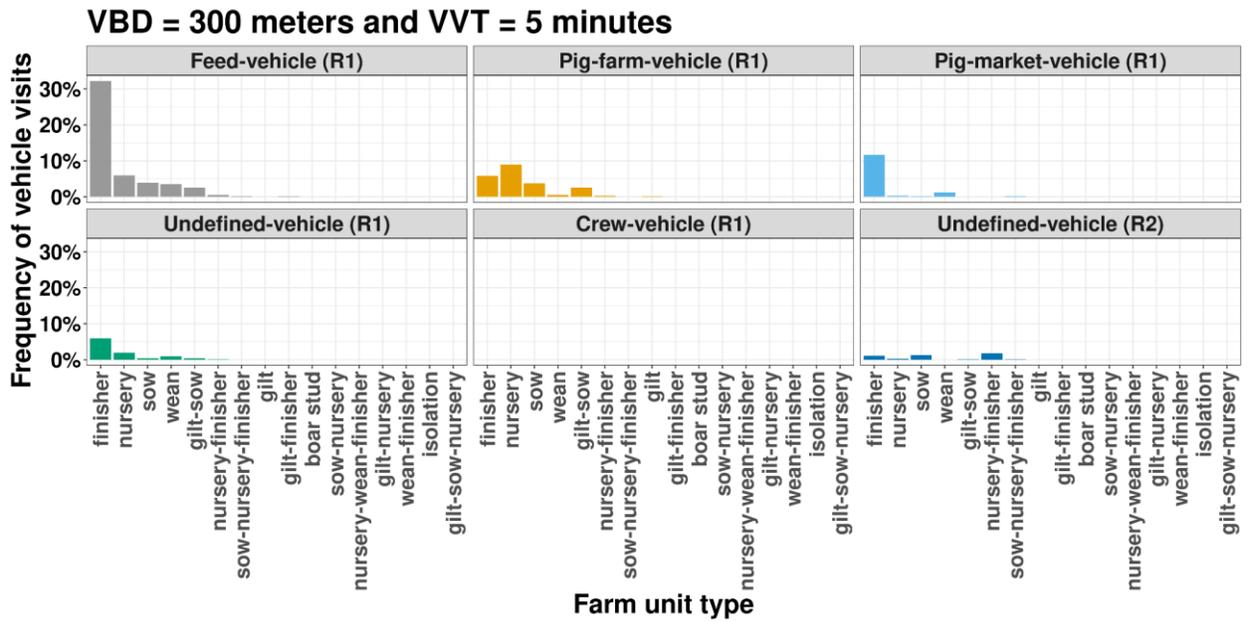

439
440
441 **Figure S10. Vehicle visit frequency company by farm type**
442
443

444

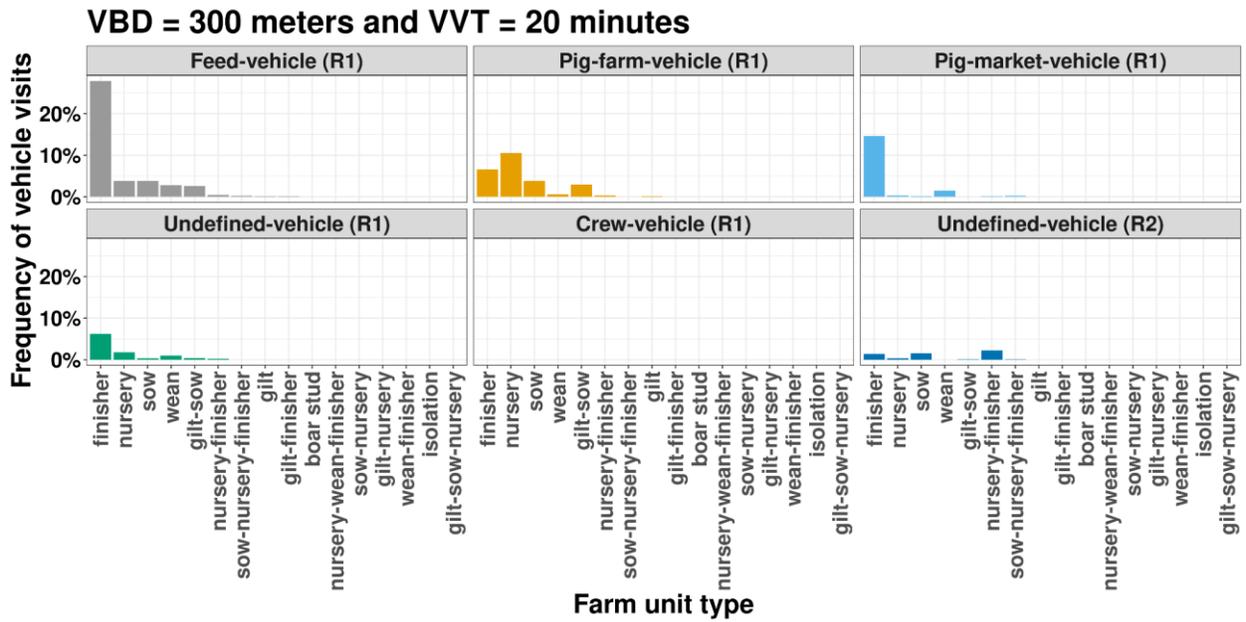

445
446 **Figure S11. Vehicle visit frequency company by farm type**
447

448

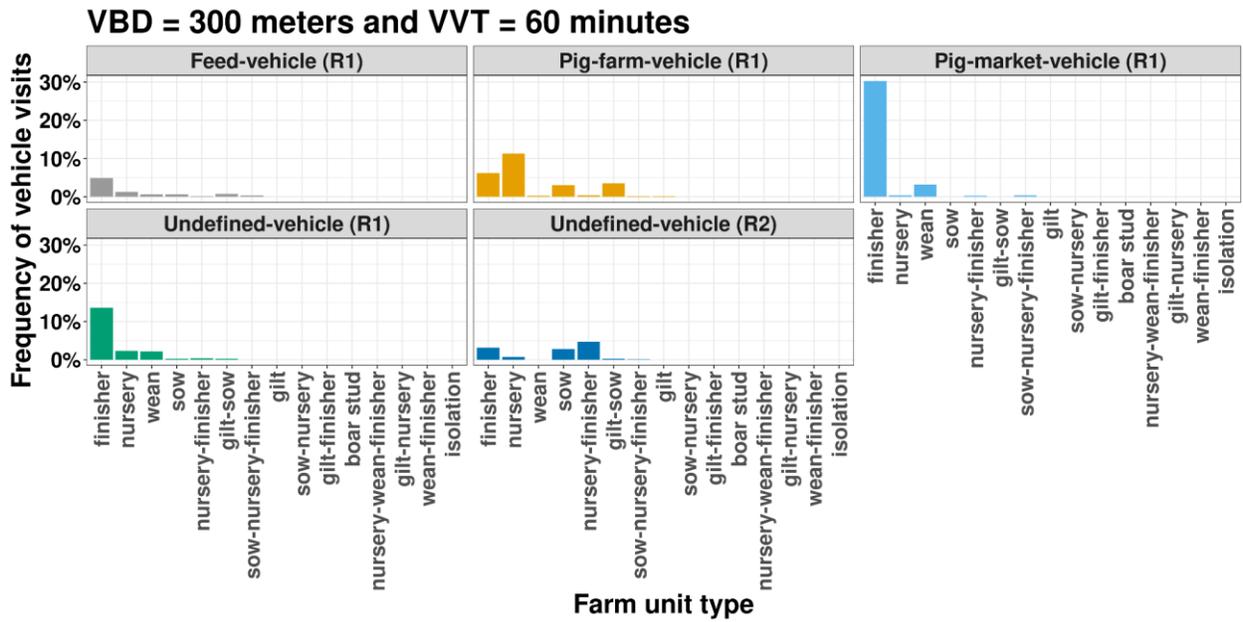

449
450 **Figure S12. Vehicle visit frequency company by farm type**
451
452
453
454

455

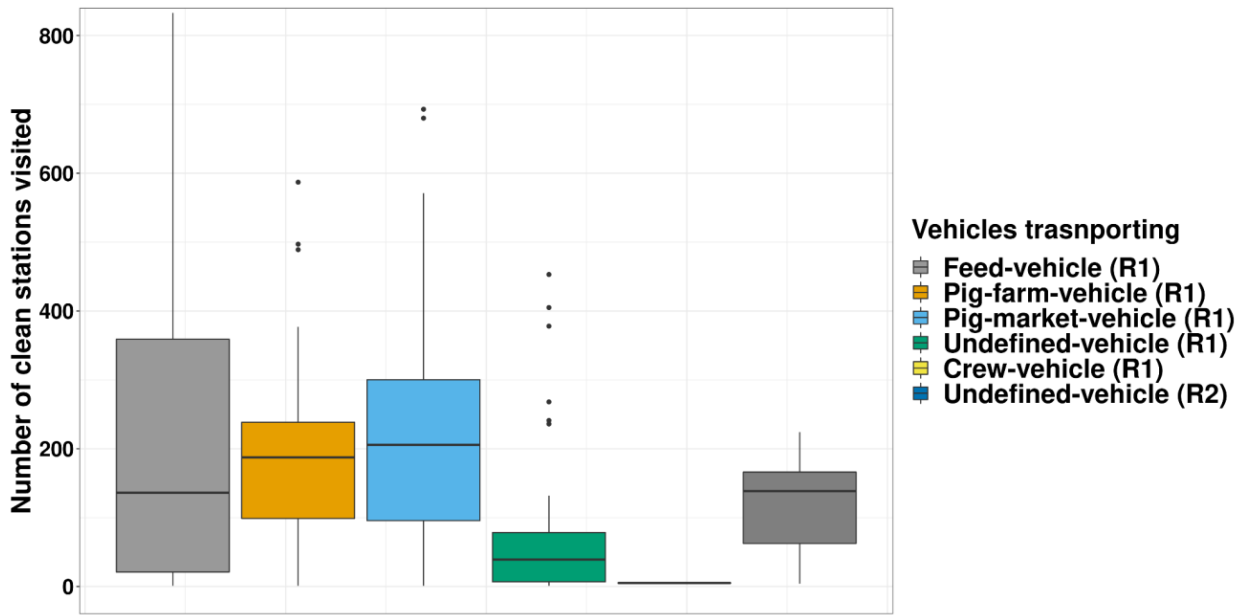

456
457 **Figure S13. The frequency of vehicle visits to cleaning stations.** The y-axis represents the number of
458 cleaning stations for each vehicle in 12 months. Vehicle types are grouped in region one (R1) and region
459 two (R2).
460
461
462
463
464
465
466
467
468

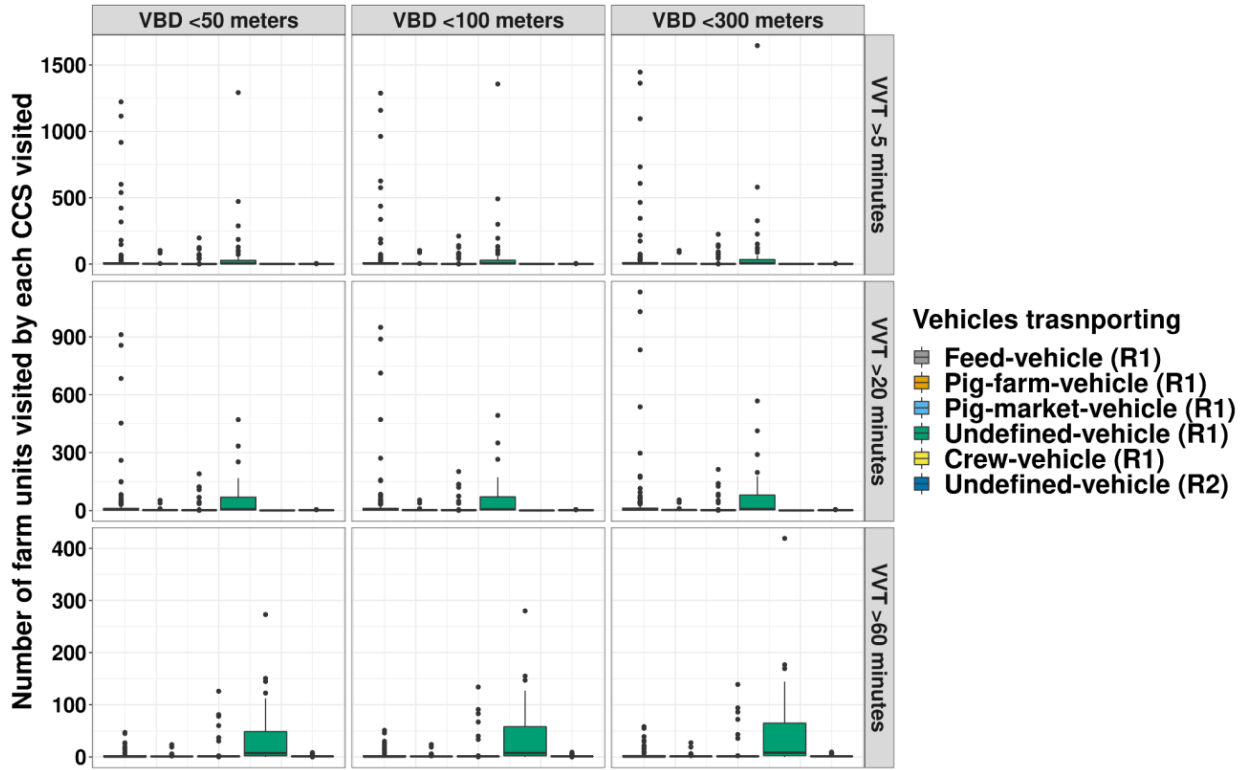

469
470
471  **Figure S14. The ratio of vehicles visiting farms and cleaning stations.** The y-axis represents the ratio
472  between the number of farm units and clean stations visited by each vehicle for one year. Vehicle
473  transporting role types in the legend are grouped in region 1 (R1) and region 2 (R2) (see Figure S15 to
474  evaluate the distribution without outliers).
475
476

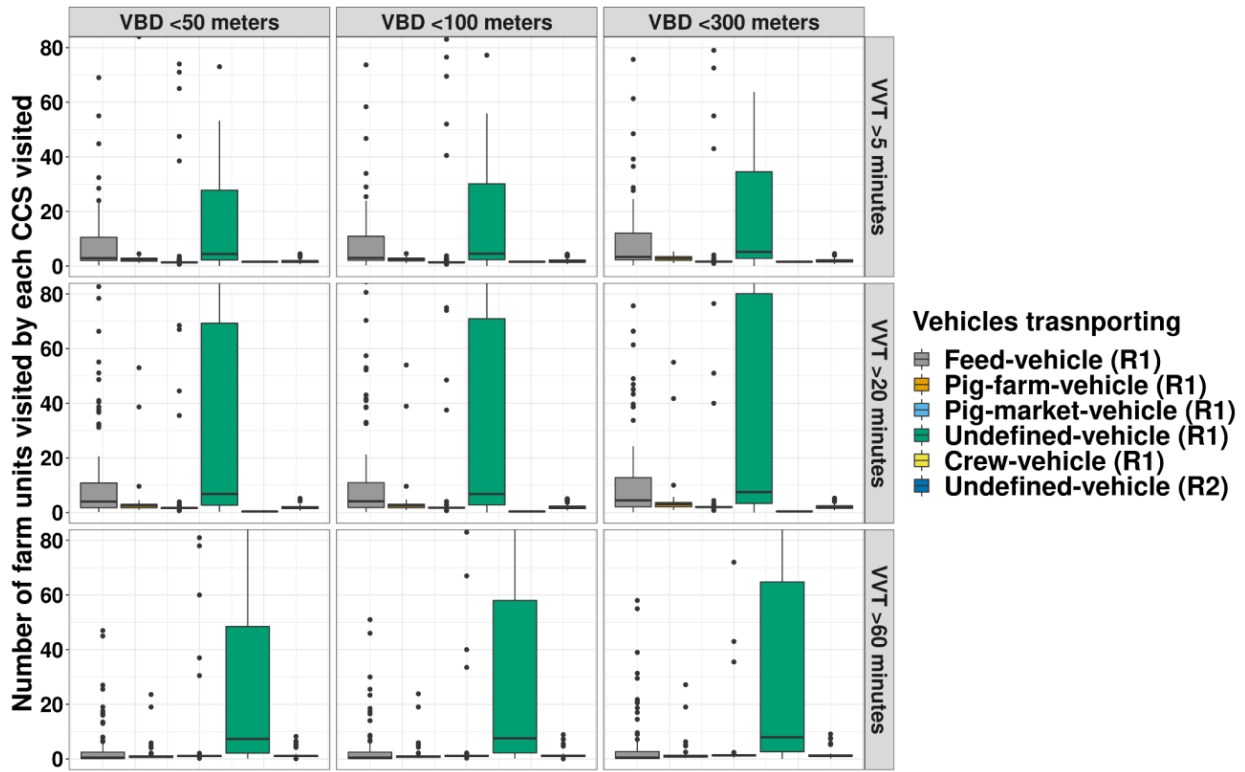

**Figure S15. Ratio of vehicles visiting farms and cleaning stations.** The y-axis represents the ratio between the number of farm units and clean stations visited by each vehicle for one year. Vehicle transporting role types in the legend are grouped in region 1 (R1) and region 2 (R2). The y-axis maximum was limited to 80 to avoid distortion created by the outliers.

484  **Table S6.** The median ratio and the interquartile range (IQR) of vehicles transporting feed visiting farm
485  units and cleaning stations.

|  | VVT | | |
| --- | --- | --- | --- |
| **VBD** | **5 minutes** | **20 minutes** | **60 minutes** |
| **50 meters** | 2.9 (2.1-10.6) (R1) | 4 (1.8-10.8) (R1) | 0.5 (0.1-2.5) (R1) |
| **100 meters** | 3 (2.1-10.9) (R1) | 4.1 (1.8-10.9) (R1) | 0.5 (0.1-2.5) (R1) |
| **300 meters** | 3.3 (2.3-12.1) (R1) | 4.4 (2-12.7) (R1) | 0.5 (0.1-2.7) (R1) |

486  (R1) = region 1; (R2) = region 2

490 **Table S7.** The median ratio and the interquartile range (IQR) of vehicles transporting pigs to farms,
491 visiting farm units, and cleaning stations.

|  | VVT | | |
|---|---|---|---|
| **VBD** | **5 minutes** | **20 minutes** | **60 minutes** |
| **50 meters** | 2.4 (1.8-2.9) (R1) | 2.5 (1.7-3) (R1) | 0.8 (0.6,1.1) (R1) |
| **100 meters** | 2.5 (1.9-3) (R1) | 2.6 (1.8-3.1) (R1) | 0.8 (0.6-1.1) (R1) |
| **300 meters** | 2.8 (2.1-3.5) (R1) | 3 (2-3.6) (R1) | 0.9 (0.6-1.3) (R1) |

492 (R1) = region 1; (R2) = region 2
493
494

495  **Table S8.** The median ratio and the interquartile range (IQR) of vehicles transporting pigs to market,
496  visiting farm units and cleaning stations.

|  | VVT | | |
| --- | --- | --- | --- |
| **VBD** | **5 minutes** | **20 minutes** | **60 minutes** |
| **50 meters** | 1.3 (1.2-1.5) (R1) | 1.7 (1.4-1.8) (R1) | 1 (0.9,1.2) (R1) |
| **100 meters** | 1.4 (1.3-1.6) (R1) | 1.7 (1.6-1.9) (R1) | 1.1 (0.9-1.3) (R1) |
| **300 meters** | 1.6 (1.5-1.8) (R1) | 2 (1.8-2.2) (R1) | 1.3 (1-1.5) (R1) |

497  (R1) = region 1; (R2) = region 2
498
499

Table S9. The median ratio and the interquartile range (IQR) of vehicles transporting crew visiting farm units and cleaning stations.

|  | VVT | | |
|---|---|---|---|
| **VBD** | **5 minutes** | **20 minutes** | **60 minutes** |
| **50 meters** | 1.6 (1.6-1.6) (R1) | 0.4 (0.4-0.4) (R1) | 0 (R1) |
| **100 meters** | 1.6 (1.6-1.6) (R1) | 0.4 (0.4-0.4) (R1) | 0 (R1) |
| **300 meters** | 1.6 (1.6-1.6) (R1) | 0.4 (0.4-0.4) (R1) | 0 (R1) |

(R1) = region 1; (R2) = region 2

507 **Table S10.** The median ratio and the interquartile range (IQR) of undefined vehicles visiting farm units
508 and cleaning stations.

|  | VVT | | |
|---|---|---|---|
| **VBD** | **5 minutes** | **20 minutes** | **60 minutes** |
| **50 meters** | 4.4 (2.2-27.7) (R1)<br>1.6 (1.3-2.1) (R2) | 6.8 (2.7-69.3) (R1)<br>1.6 (1.4-2.1) (R2) | 7.3 (2.1-48.4) (R1)<br>1 (0.8-1.3) (R2) |
| **100 meters** | 4.6 (2.4-30.2) (R1)<br>1.6 (1.4-2.2) (R2) | 6.7 (2.7-71) (R1)<br>1.7 (1.4-2.3) (R2) | 7.6 (2.2-58) (R1)<br>1.1 (0.9-1.4) (R2) |
| **300 meters** | 5.2 (2.8-34.6) (R1)<br>1.7 (1.5-2.3) (R2) | 7.5 (3.3-80.1) (R1)<br>1.8 (1.5-2.5) (R2) | 7.9 (2.6-64.8) (R1)<br>1.2 (0.9-1.4) (R2) |

509 (R1) = region 1; (R2) = region 2

## Section 3. Vehicle network metrics

512 **Table S11. Network metric of vehicle movements networks.** Values represent the median and the interquartile range (IQR).

| Metric | $d$ (%) | Combined vehicles (R1) | Feed-vehicle (R1) | Pig-vehicle (R1) | Market-vehicle (R1) | Undefined (R1) | Crew-vehicle (R1) | Undefined (R2) |
|---|---|---|---|---|---|---|---|---|
| **Nodes** | 0 | 2,159 (2,159-2,159) | 2,151 (2,151-2,151) | 2,103 (2,103-2,103) | 1,618 (1,618-1,618) | 1,848 (1,848-1,848) | 7 (7-7) | 450 (450-450) |
| | 10 | 2,159 (2,159-2,159) | 2,151 (2,151-2,151) | 2,103 (2,103-2,103) | 1,618 (1,617-1,618) | 1,848 (1,848-1,848) | 7 (7-7) | 450 (450-450) |
| | 50 | 2,159 (2,158-2,159) | 2,151 (2,151-2,151) | 2,103 (2,103-2,103) | 1,617 (1,616-1,617) | 1,848 (1,847-1,848) | 7 (7-7) | 449 (448-449) |
| | 80 | 2,158 (2,158-2,159) | 2,151 (2,151-2,151) | 2,102 (2,102-2,103) | 1,612 (1,612-1,615) | 1,847 (1,847-1,847) | 7 (6-7) | 447 (446-447) |
| | 90 | 2,158 (2,158-2,158) | 2,151 (2,151-2,151) | 2,102 (2,101-2,102) | 1,607 (1,605-1,609) | 1,847 (1,847-1,848) | 6 (6-7) | 444 (444-444) |
| | 100 | 2,158 (2,158-2,158) | 2,151 (2,151-2,151) | 2,101 (2,101-2,101) | 1,593 (1,593-1,593) | 1,847 (1,847-1,847) | 6 (6-6) | 442 (442-442) |
| **Edge static** | 0 | 1,232,684 (1,232,684-1,232,684) | 1,018,941 (1,018,941-1,018,941) | 207,232 (207,232-207,232) | 139,786 (139,786-139,786) | 290,960 (290,960-290,960) | 23 (23-23) | 21,385 (21,385-21,385) |
| | 10 | 1,161,600 (1,161,342-1,161,990) | 963,416 (962,678-963,797) | 185,370 (184,994-185,524) | 116,088 (115,991-116,260) | 271,140 (270,501-271,390) | 20 (18-21) | 19,001 (18,960-19,128) |
| | 50 | 954,620 (954,306-954,916) | 812,846 (812,615-813,236) | 106,854 (106,641-107,014) | 53,727 (53,550-53,840) | 227,676 (227,412-228,421) | 13 (11-14) | 11,198 (11,119-11,296) |
|  | 80 | 850,697 (850,524-850,832) | 741,734 (741,329-742,119) | 61,246 (61,053-61,433) | 28,792 (28,683-29,001) | 211,748 (211,663-212,058) | 8 (7-10) | 6,569 (6,546-6,623) |
|  | 90 | 821,368 (821,309-821,421) | 721,940 (721,715-722,116) | 47,938 (47,772-47,995) | 22,482 (22,452-22,668) | 208,183 (208,114-208,308) | 8 (7-8) | 5,196 (5,144-5,212) |
|  | 100 | 793,827 (793,827-793,827) | 703,726 (703,726-703,726) | 34,769 (34,769-34,769) | 16,973 (16,973-16,973) | 204,761 (204,761-204,761) | 7 (7-7) | 3,857 (3,857-3,857) |
| **Edge temporal** | 0 | 5,583,703 (5,583,703-5,583,703) | 3,846,333 (3,846,333-3,846,333) | 841,987 (841,987-841,987) | 359,796 (359,796-359,796) | 535,563 (535,563-535,563) | 24 (24-24) | 128,483 (128,483-128,483) |
|  | 10 | 4,675,428 (4,674,189-4,676,232) | 3,249,931 (3,248,909-3,251,477) | 669,330 (668,443-670,426) | 263,362 (262,922-263,882) | 491,788 (490,917-492,270) | 21 (18-22) | 93,402 (93,056-95,064) |
|  | 50 | 3,058,733 (3,057,575-3,059,771) | 2,242,121 (2,241,763-2,243,493) | 309,621 (309,132-310,337) | 96,466 (96,305-96,900) | 409,836 (409,451-410,717) | 13 (12-14) | 36,154 (35,849-36,274) |
|  | 80 | 2,574,086 (2,573,529-2,574,522) | 1,951,840 (1,950,915-1,952,681) | 186,946 (186,574-187,086) | 51,218 (51,010-51,384) | 383,864 (383,654-384,260) | 8 (7-10) | 20,406 (20,335-20,542) |
|  | 90 | 2,463,853 (2,463,306-2,464,496) | 1,885,504 (1,884,857-1,885,864) | 158,806 (158,600-158,871) | 41,410 (41,314-41,635) | 378,255 (378,212-378,508) | 8 (7-8) | 17,023 (16,952-17,089) |
|  | 100 | 2,370,612 (2,370,612-2,370,612) | 1,828,595 (1,828,595-1,828,595) | 135,173 (135,173-135,173) | 33,520 (33,520-33,520) | 373,317 (373,317-373,317) | 7 (7-7) | 14,336 (14,336-14,336) |
| **Density** | 0 | 0.194343 | 0.160644 | 0.032672 | 0.022038 | 0.045872 | 4e-06 (4e-06- | 0.001003 |
|  |  | (0.194343-0.194343) | (0.160644-0.160644) | (0.032672-0.032672) | (0.022038-0.022038) | (0.045872-0.045872) | 4e-06) | (0.001003-0.001003) |
|  | 10 | 0.183136 (0.183095-0.183197) | 0.15189 (0.151774-0.15195) | 0.029225 (0.029166-0.029249) | 0.018302 (0.018287-0.018329) | 0.042747 (0.042647-0.042787) | 3e-06 (3e-06-3e-06) | 0.000891 (0.000889-0.000897) |
|  | 50 | 0.150504 (0.150454-0.15055) | 0.128152 (0.128115-0.128213) | 0.016846 (0.016813-0.016872) | 0.00847 (0.008443-0.008488) | 0.035895 (0.035853-0.036012) | 2e-06 (2e-06-2e-06) | 0.000525 (0.000521-0.00053) |
|  | 80 | 0.134119 (0.134092-0.13414) | 0.11694 (0.116876-0.117001) | 0.009656 (0.009625-0.009685) | 0.004539 (0.004522-0.004572) | 0.033384 (0.03337-0.033433) | 1e-06 (1e-06-2e-06) | 0.000308 (0.000307-0.00031) |
|  | 90 | 0.129495 (0.129486-0.129504) | 0.11382 (0.113784-0.113847) | 0.007558 (0.007532-0.007567) | 0.003544 (0.00354-0.003574) | 0.032822 (0.032811-0.032841) | 1e-06 (1e-06-1e-06) | 0.000244 (0.000241-0.000244) |
|  | 100 | 0.125153 (0.125153-0.125153) | 0.110948 (0.110948-0.110948) | 0.005482 (0.005482-0.005482) | 0.002676 (0.002676-0.002676) | 0.032282 (0.032282-0.032282) | 1e-06 (1e-06-1e-06) | 0.000181 (0.000181-0.000181) |
| In-degree | 0 | 476 (476-476) | 338 (338-338) | 63 (63-63) | 38 (38-38) | 100 (100-100) | 0.0 (0.0-0.0) | 0.0 (0.0-0.0) |
|  | 10 | 447 (446-449) | 327 (326-327) | 55 (55-56) | 31 (31-31) | 91 (90-91) | 0.0 (0.0-0.0) | 0.0 (0.0-0.0) |
|  | 50 | 370 (369-372) | 288 (288-288) | 30 (30-30) | 14 (14-14) | 63 (63-64) | 0.0 (0.0-0.0) | 0.0 (0.0-0.0) |
|  | 80 | 329 (329-330) | 259 (259-260) | 17 (17-17) | 7 (7-7) | 51 (51-52) | 0.0 (0.0-0.0) | 0.0 (0.0-0.0) |
|  | 90 | 318 (317-319) | 250 (249-250) | 14 (14-14) | 5 (5-5) | 48 (48-49) | 0.0 (0.0-0.0) | 0.0 (0.0-0.0) |
| | 100 | 306 (306-306) | 243 (243-243) | 10 (10-10) | 3 (3-3) | 46 (46-46) | 0.0 (0.0-0.0) | 0.0 (0.0-0.0) |
| **Out-degree** | 0 | 477 (477-477) | 336 (336-336) | 61 (61-61) | 38 (38-38) | 100 (100-100) | 0.0 (0.0-0.0) | 0.0 (0.0-0.0) |
| | 10 | 448 (447-449) | 327 (326-328) | 54 (54-55) | 32 (31-32) | 90 (89-90) | 0.0 (0.0-0.0) | 0.0 (0.0-0.0) |
| | 50 | 368 (367-369) | 286 (286-287) | 28 (28-28) | 13 (13-13) | 63 (63-64) | 0.0 (0.0-0.0) | 0.0 (0.0-0.0) |
| | 80 | 326 (325-326) | 257 (257-258) | 14 (14-15) | 7 (7-7) | 52 (51-52) | 0.0 (0.0-0.0) | 0.0 (0.0-0.0) |
| | 90 | 313 (312-314) | 248 (247-249) | 11 (11-11) | 5 (5-5) | 49 (49-50) | 0.0 (0.0-0.0) | 0.0 (0.0-0.0) |
| | 100 | 299 (299-299) | 239 (239-239) | 7 (7-7) | 3 (3-3) | 46 (46-46) | 0.0 (0.0-0.0) | 0.0 (0.0-0.0) |
| **Degree centrality** | 0 | 0.29823 (0.29823-0.29823) | 0.311481 (0.311481-0.311481) | 0.136366 (0.136366-0.136366) | 0.116809 (0.116809-0.116809) | 0.178384 (0.178384-0.178384) | 0.001586 (0.001586-0.001586) | 0.028995 (0.028995-0.028995) |
| | 10 | 0.299724 (0.29933-0.300038) | 0.311474 (0.310933-0.312744) | 0.127608 (0.126761-0.129057) | 0.099483 (0.098516-0.100368) | 0.162706 (0.161645-0.165415) | 0.001586 (0.001189-0.001586) | 0.026187 (0.025914-0.026745) |
| | 50 | 0.296262 (0.29459-0.297385) | 0.299837 (0.299194-0.301549) | 0.089664 (0.087329-0.090554) | 0.052727 (0.05174-0.053753) | 0.130296 (0.128948-0.132017) | 0.00119 (0.001189-0.00119) | 0.017398 (0.017056-0.017565) |
| | 80 | 0.297509 (0.296628-0.298509) | 0.293071 (0.292608-0.293606) | 0.061436 (0.060478-0.062876) | 0.03113 (0.030864-0.031568) | 0.125307 (0.124367-0.125836) | 0.000793 (0.000595-0.000793) | 0.012739 (0.012574-0.013094) |
|  | 90 | 0.299853 (0.297737-0.300132) | 0.294164 (0.292948-0.294596) | 0.053473 (0.052697-0.054221) | 0.024375 (0.022926-0.026701) | 0.124773 (0.123799-0.125041) | 0.000595 (0.000595-0.000595) | 0.012156 (0.012101-0.012402) |
|  | 100 | 0.299708 (0.299708-0.299708) | 0.293855 (0.293855-0.293855) | 0.043583 (0.043583-0.043583) | 0.020962 (0.020962-0.020962) | 0.123645 (0.123645-0.123645) | 0.000595 (0.000595-0.000595) | 0.01184 (0.01184-0.01184) |
| **Betweenness centrality** | 0 | 0.05903 (0.05903-0.05903) | 0.066982 (0.066982-0.066982) | 0.091472 (0.091472-0.091472) | 0.047358 (0.047358-0.047358) | 0.025259 (0.025259-0.025259) | 0.0 (0.0-0.0) | 0.000964 (0.000964-0.000964) |
|  | 10 | 0.061168 (0.060929-0.061632) | 0.063479 (0.062075-0.063886) | 0.091433 (0.09141-0.091454) | 0.045614 (0.044915-0.046598) | 0.025125 (0.024426-0.025995) | 0.0 (0.0-0.0) | 0.000977 (0.000927-0.001003) |
|  | 50 | 0.064494 (0.062939-0.064879) | 0.062042 (0.060605-0.064429) | 0.091931 (0.091359-0.091938) | 0.047957 (0.045256-0.051347) | 0.023962 (0.023665-0.02436) | 1e-06 (0.0-1e-06) | 0.000814 (0.000694-0.000992) |
|  | 80 | 0.063994 (0.063659-0.065105) | 0.063281 (0.062188-0.064398) | 0.091811 (0.091795-0.091929) | 0.048833 (0.044646-0.051741) | 0.025787 (0.024456-0.026767) | 0.0 (0.0-0.0) | 0.000815 (0.000766-0.000983) |
|  | 90 | 0.064173 (0.063734-0.064689) | 0.063944 (0.063044-0.064587) | 0.091753 (0.091659-0.091801) | 0.048487 (0.046355-0.051334) | 0.024439 (0.023612-0.025553) | 0.0 (0.0-0.0) | 0.001031 (0.000905-0.001251) |
|  | 100 | 0.062491 (0.062491-0.062491) | 0.064005 (0.064005-0.064005) | 0.091383 (0.091383-0.091383) | 0.043759 (0.043759-0.043759) | 0.023805 (0.023805-0.023805) | 0.0 (0.0-0.0) | 0.001697 (0.001697-0.001697) |

**Table S12.** Cumulative reduction of network metrics as cleaning effectiveness (*d*) increase

| Metric | *d* (%) | Combined | Feed-vehicle (R1) (%) | Pig-vehicle (R1) (%) | Market-vehicle (R1) | Undefined (R1) (%) | Crew-vehicle (R1) | Undefined (R2) (%) |
|---|---|---|---|---|---|---|---|---|

|  |  | vehicles (R1) (%) |  |  | (%) |  | (%) |  |
|---|---|---|---|---|---|---|---|---|
| **Nodes** | **0** | 0 | 0 | 0 | 0 | 0 | 0 | 0 |
|  | **10** | 0 | 0 | 0 | 0 | 0 | 0 | 0 |
|  | **50** | 0 | 0 | 0 | 0.06 | 0.03 | 0 | 0.22 |
|  | **80** | 0.02 | 0 | 0.02 | 0.34 | 0.05 | 0 | 0.67 |
|  | **90** | 0.05 | 0 | 0.07 | 0.68 | 0.05 | 7.14 | 1.33 |
|  | **100** | 0.05 | 0 | 0.1 | 1.55 | 0.05 | 14.29 | 1.78 |
| **Edge static** | **0** | 0 | 0 | 0 | 0 | 0 | 0 | 0 |
|  | **10** | 5.77 | 5.45 | 10.55 | 16.95 | 6.81 | 10.87 | 11.15 |
|  | **50** | 22.56 | 20.23 | 48.44 | 61.56 | 21.75 | 43.48 | 47.64 |
|  | **80** | 30.99 | 27.21 | 70.45 | 79.4 | 27.22 | 63.04 | 69.28 |
|  | **90** | 33.37 | 29.15 | 76.87 | 83.92 | 28.45 | 67.39 | 75.7 |
|  | **100** | 35.6 | 30.94 | 83.22 | 87.86 | 29.63 | 69.57 | 81.96 |
| **Edge temporal** | **0** | 0 | 0 | 0 | 0 | 0 | 0 | 0 |
|  | **10** | 16.27 | 15.51 | 20.51 | 26.8 | 8.17 | 12.5 | 27.3 |
|  | 50 | 45.22 | 41.71 | 63.23 | 73.19 | 23.48 | 45.83 | 71.86 |
|  | 80 | 53.9 | 49.25 | 77.8 | 85.76 | 28.33 | 64.58 | 84.12 |
|  | 90 | 55.87 | 50.98 | 81.14 | 88.49 | 29.37 | 68.75 | 86.75 |
|  | 100 | 57.54 | 52.46 | 83.95 | 90.68 | 30.29 | 70.83 | 88.84 |
| Density | 0 | 0 | 0 | 0 | 0 | 0 | 0 | 0 |
|  | 10 | 5.76 | 5.45 | 10.53 | 16.97 | 6.8 | 0 | 11 |
|  | 50 | 22.56 | 20.23 | 48.42 | 61.57 | 21.76 | 0 | 48 |
|  | 80 | 30.99 | 27.2 | 70.43 | 79.4 | 27.23 | 0 | 69 |
|  | 90 | 33.36 | 29.15 | 76.86 | 83.94 | 28.45 | 0 | 76 |
|  | 100 | 35.6 | 30.93 | 83.23 | 87.84 | 29.63 | 0 | 82 |
| In-degree | 0 | 0 | 0 | 0 | 0 | 0 | 0 | 0 |
|  | 10 | 6.09 | 3.25 | 12.7 | 18.42 | 9 | 0 | 0 |
|  | 50 | 22.27 | 14.79 | 52.38 | 63.16 | 37 | 0 | 0 |
|  | 80 | 30.88 | 23.37 | 73.02 | 81.58 | 49 | 0 | 0 |
|  | 90 | 33.19 | 26.04 | 77.78 | 86.84 | 52 | 0 | 0 |
| | 100 | 35.71 | 28.11 | 84.13 | 92.11 | 54 | 0 | 0 |
| **Out-degree** | 0 | 0 | 0 | 0 | 0 | 0 | 0 | 0 |
| | 10 | 6.08 | 2.68 | 10.66 | 15.79 | 10 | 0 | 0 |
| | 50 | 22.85 | 14.73 | 54.1 | 65.79 | 37 | 0 | 0 |
| | 80 | 31.66 | 23.51 | 77.05 | 81.58 | 48 | 0 | 0 |
| | 90 | 34.38 | 26.19 | 81.97 | 86.84 | 51 | 0 | 0 |
| | 100 | 37.32 | 28.87 | 88.52 | 92.11 | 54 | 0 | 0 |

**Table S13.** Proportion of edges in each pathogen stability category.

| d (%) | Pathogen stability | Combined vehicles (R1) | Feed-vehicle (R1) | Pig-vehicle (R1) | Market-vehicle (R1) | Undefined (R1) | Crew-vehicle (R1) | Undefined (R2) |
|---|---|---|---|---|---|---|---|---|
| 0 | >0.8-1 | 6.080015 | 5.172381 | 10.40432 | 5.621519 | 6.107031 | 29.16667 | 5.427177 |
| | >0.6-≤0.8 | 4.938855 | 5.197704 | 3.899585 | 4.738241 | 4.848729 | 0 | 4.358553 |
| | >0.4-≤0.6 | 6.20907 | 6.588587 | 4.446981 | 6.002568 | 6.392712 | 0 | 4.690893 |
|   | >0.2-≤0.4 | 10.57021 | 10.2572 | 12.05648 | 11.02013 | 10.17901 | 16.66667 | 10.89249 |
|   | >0-≤0.2 | 72.20185 | 72.78413 | 69.19264 | 72.61754 | 72.47252 | 54.16667 | 74.63089 |
| 10 | >0.8-1 | 7.213415 | 6.06979 | 13.07741 | 7.553102 | 6.626744 | 33.33333 | 7.45063 |
|   | >0.6-≤0.8 | 5.678048 | 5.93168 | 4.692528 | 5.992702 | 5.178456 | 0 | 5.620359 |
|   | >0.4-≤0.6 | 6.927741 | 7.331079 | 4.999029 | 7.2201 | 6.746003 | 0 | 5.768109 |
|   | >0.2-≤0.4 | 11.45075 | 11.09716 | 13.33281 | 12.78468 | 10.56605 | 19.04762 | 13.02067 |
|   | >0-≤0.2 | 68.71791 | 69.57065 | 63.87679 | 66.45492 | 70.85204 | 52.38095 | 68.27674 |
| 50 | >0.8-1 | 10.73732 | 8.494457 | 28.18026 | 19.23848 | 7.838877 | 53.84615 | 18.98657 |
|   | >0.6-≤0.8 | 7.405305 | 7.418155 | 8.3037 | 11.4896 | 5.726688 | 0 | 10.60587 |
|   | >0.4-≤0.6 | 8.091014 | 8.421223 | 6.006214 | 10.79142 | 7.216798 | 0 | 8.716757 |
|   | >0.2-≤0.4 | 12.17055 | 11.7877 | 15.29612 | 16.75055 | 10.84411 | 15.38462 | 17.32703 |
|   | >0-≤0.2 | 61.60976 | 63.89822 | 42.23066 | 41.91084 | 68.3889 | 23.07692 | 44.39005 |
| 80 | >0.8-1 | 12.5005 | 9.499423 | 46.55355 | 34.32354 | 8.27766 | 82.35294 | 33.37907 |
|   | >0.6-≤0.8 | 7.766019 | 7.600113 | 11.54879 | 15.42411 | 5.738092 | 0 | 13.93674 |
| | >0.4-≤0.6 | 7.850359 | 8.205772 | 4.873332 | 10.64264 | 7.104721 | 0 | 8.825619 |
| | >0.2-≤0.4 | 11.16 | 11.16921 | 11.50011 | 14.37957 | 10.53158 | 5.882353 | 15.42891 |
| | >0-≤0.2 | 60.71817 | 63.52438 | 25.56567 | 25.14716 | 68.36787 | 11.76471 | 28.69184 |
| 90 | >0.8-1 | 12.97648 | 9.75047 | 54.75598 | 41.71889 | 8.373981 | 93.33333 | 39.91365 |
| | >0.6-≤0.8 | 7.768909 | 7.571424 | 12.72941 | 16.526 | 5.723388 | 0 | 14.7242 |
| | >0.4-≤0.6 | 7.651857 | 8.055459 | 3.787022 | 9.700438 | 7.017224 | 0 | 8.597192 |
| | >0.2-≤0.4 | 10.7101 | 10.91568 | 8.543155 | 11.95953 | 10.43331 | 0 | 13.21154 |
| | >0-≤0.2 | 60.8946 | 63.69504 | 20.19326 | 20.21468 | 68.45091 | 0 | 23.59455 |
| 100 | >0.8-1 | 13.39553 | 9.961583 | 64.27763 | 50.55191 | 8.454209 | 100 | 47.25865 |
| | >0.6-≤0.8 | 7.725769 | 7.50456 | 13.93991 | 17.5537 | 5.676945 | 0 | 15.30413 |
| | >0.4-≤0.6 | 7.418633 | 7.879875 | 2.306674 | 8.111575 | 6.94825 | 0 | 7.896205 |
| | >0.2-≤0.4 | 10.21297 | 10.6555 | 4.430619 | 8.317422 | 10.30947 | 0 | 9.946987 |
| | >0-≤0.2 | 61.2471 | 63.99848 | 15.04516 | 15.46539 | 68.61113 | 0 | 19.59403 |

519
520

521 **Table S14.** The the median and the interquartile range (IQR) of edges according the pathogen stability categories.

| Pathogen stability | $d$ (%) | Combined vehicles (R1) | Feed-vehicle (R1) | Pig-vehicle (R1) | Market-vehicle (R1) | Undefined (R1) | Crew-vehicle (R1) | Undefined (R2) |
|---|---|---|---|---|---|---|---|---|
| >0.8 - 1 | 0 | 339,490 (339,490-339,490) | 198,947 (198,947-198,947) | 87,603 (87,603-87,603) | 20,226 (20,226-20,226) | 32,707 (32,707-32,707) | 7 (7-7) | 6,973 (6,973-6,973) |
| | 10 | 337,258 (337,228-337,288) | 197,264 (197,191-197,284) | 87,531 (87,530-87,538) | 19,892 (19,878-19,898) | 32,590 (32,574-32,598) | 7 (7-7) | 6,959 (6,954-6,965) |
| | 50 | 328,426 (328,335-328,499) | 190,456 (190,379-190,520) | 87,252 (87,236-87,272) | 18,558 (18,538-18,588) | 32,126 (32,101-32,162) | 7 (7-7) | 6,864 (6,857-6,881) |
| | 80 | 321,774 (321,749-321,895) | 185,414 (185,324-185,481) | 87,030 (87,006-87,045) | 17,580 (17,577-17,602) | 31,775 (31,766-31,794) | 7 (7-7) | 6,812 (6,807-6,826) |
| | 90 | 319,722 (319,622-319,815) | 183,846 (183,734-183,882) | 86,956 (86,940-86,967) | 17,276 (17,261-17,282) | 31,675 (31,666-31,683) | 7 (7-7) | 6,794 (6,792-6,798) |
| | 100 | 317,556 (317,556-317,556) | 182,157 (182,157-182,157) | 86,886 (86,886-86,886) | 16,945 (16,945-16,945) | 31,561 (31,561-31,561) | 7 (7-7) | 6,775 (6,775-6,775) |
| >0.6- ≤0.8 | 0 | 275,771 (275,771-275,771) | 199,921 (199,921-199,921) | 32,834 (32,834-32,834) | 17,048 (17,048-17,048) | 25,968 (25,968-25,968) | 0.0 (0.0-0.0) | 5,600 (5,600-5,600) |
| | 10 | 265,473 (265,394-265,625) | 192,776 (192,684-192,930) | 31,408 (31,362-31,450) | 15,782 (15,758-15,837) | 25,467 (25,434-25,516) | 0.0 (0.0-0.0) | 5,250 (5,216-5,309) |
| | 50 | 226,508 (226,311-226,641) | 166,324 (166,042-166,448) | 25,710 (25,544-25,748) | 11,084 (10,984-11,125) | 23,470 (23,436-23,584) | 0.0 (0.0-0.0) | 3,834 (3,822-3,890) |
| | 80 | 199,904 (199,761-199,967) | 148,342 (148,158-148,510) | 21,590 (21,537-21,637) | 7,900 (7,838-7,980) | 22,026 (22,008-22,063) | 0.0 (0.0-0.0) | 2,844 (2,828-2,878) |
| | 90 | 191,414 (191,382-191,530) | 142,760 (142,696-142,847) | 20,215 (20,144-20,256) | 6,844 (6,819-6,900) | 21,649 (21,593-21,668) | 0.0 (0.0-0.0) | 2,506 (2,490-2,533) |
| | 100 | 183,148 (183,148-183,148) | 137,228 (137,228-137,228) | 18,843 (18,843-18,843) | 5,884 (5,884-5,884) | 21,193 (21,193-21,193) | 0.0 (0.0-0.0) | 2,194 (2,194-2,194) |
| >0.4-≤0.6 | 0 | 346,696 (346,696-346,696) | 253,419 (253,419-253,419) | 37,443 (37,443-37,443) | 21,597 (21,597-21,597) | 34,237 (34,237-34,237) | 0.0 (0.0-0.0) | 6,027 (6,027-6,027) |
| | 10 | 323,902 (323,806-324,080) | 238,255 (238,178-238,433) | 33,460 (33,397-33,544) | 19,015 (18,986-19,048) | 33,176 (33,151-33,228) | 0.0 (0.0-0.0) | 5,388 (5,313-5,476) |
| | 50 | 247,482 (247,090-247,742) | 188,814 (188,631-188,972) | 18,596 (18,429-18,809) | 10,410 (10,400-10,454) | 29,577 (29,445-29,640) | 0.0 (0.0-0.0) | 3,152 (3,113-3,179) |
| | 80 | 202,075 (201,815-202,099) | 160,164 (160,102-160,312) | 9,110 (8,988-9,173) | 5,451 (5,420-5,501) | 27,272 (27,206-27,288) | 0.0 (0.0-0.0) | 1,801 (1,762-1,826) |
| | 90 | 188,530 (188,346-188,684) | 151,886 (151,770-152,052) | 6,014 (5,992-6,070) | 4,017 (3,962-4,060) | 26,543 (26,531-26,606) | 0.0 (0.0-0.0) | 1,464 (1,420-1,508) |
| | 100 | 175,867 | 144,091 | 3,118 (3,118- | 2,719 (2,719- | 25,939 (25,939- | 0.0 (0.0- | 1,132 (1,132-1,132) |

| | | | | | | | |
|---|---|---|---|---|---|---|---|
| | | (175,867-175,867) | (144,091-144,091) | 3,118) | 2,719) | 25,939) | 0.0) | |
| >0.2-≤0.4 | 0 | 590,209 (590,209-590,209) | 394,526 (394,526-394,526) | 101,514 (101,514-101,514) | 39,650 (39,650-39,650) | 54,515 (54,515-54,515) | 4 (4-4) | 13,995 (13,995-13,995) |
| | 10 | 535,372 (535,231-536,128) | 360,650 (360,462-361,010) | 89,240 (89,206-89,287) | 33,670 (33,638-33,747) | 51,962 (51,811-52,014) | 4 (4-4) | 12,162 (12,136-12,332) |
| | 50 | 372,264 (372,054-372,582) | 264,294 (264,080-264,496) | 47,360 (47,249-47,546) | 16,158 (16,120-16,234) | 44,443 (44,403-44,463) | 2 (0.25-4) | 6,264 (6,192-6,350) |
| | 80 | 287,268 (287,122-287,510) | 218,005 (217,780-218,165) | 21,499 (21,190-21,561) | 7,365 (7,318-7,408) | 40,427 (40,368-40,528) | 0.5 (0.00-1.00) | 3,148 (3,128-3,184) |
| | 90 | 263,881 (263,674-264,004) | 205,816 (205,679-205,881) | 13,567 (13,495-13,664) | 4,952 (4,929-5,020) | 39,464 (39,448-39,557) | 0.0 (0.00-1.00) | 2,249 (2,204-2,276) |
| | 100 | 242,110 (242,110-242,110) | 194,846 (194,846-194,846) | 5,989 (5,989-5,989) | 2,788 (2,788-2,788) | 38,487 (38,487-38,487) | 0.0 (0.00-0.00) | 1,426 (1,426-1,426) |
| >0-≤0.2 | 0 | 4,031,537 (4,031,537-4,031,537) | 2,799,520 (2,799,520-2,799,520) | 582,593 (582,593-582,593) | 261,275 (261,275-261,275) | 388,136 (388,136-388,136) | 13 (13-13) | 95,888 (95,888-95,888) |
| | 10 | 3,212,856 (3,212,044-3,214,328) | 2,260,998 (2,259,725-2,263,331) | 427,546 (427,038-428,573) | 175,017 (174,582-175,528) | 348,442 (348,056-348,992) | 11 (8-12) | 63,772 (63,416-65,052) |
| | 50 | 1,884,478 (1,883,543- | 1,432,676 (1,431,989- | 130,755 (130,530- | 40,430 (40,255- | 280,282 (279,878- | 3 (3-6) | 16,049 (15,791-16,105) |

|  |  |  |  |  |  |  |  |
|---|---|---|---|---|---|---|---|
|  |  | 1,885,324) | 1,433,665) | 131,004) | 40,634) | 280,767) |  |  |
|  | **80** | 1,562,938 (1,562,640-1,563,505) | 1,239,894 (1,239,519-1,240,362) | 47,794 ( 47,495-47,875) | 12,880 ( 12,791-13,000) | 262,440 (262,248-262,625) | 1.0 ( 0.00-2) | 5,855 ( 5,737- 5,884) |
|  | **90** | 1,500,354 (1,499,898-1,501,078) | 1,200,972 (1,200,786-1,201,812) | 32,068 ( 31,878-32,154) | 8,371 ( 8,308- 8,425) | 258,919 (258,798-259,050) | 0.0 ( 0.00-0.00) | 4,016 ( 3,967- 4,050) |
|  | **100** | 1,451,931 (1,451,931-1,451,931) | 1,170,273 (1,170,273-1,170,273) | 20,337 ( 20,337-20,337) | 5,184 ( 5,184- 5,184) | 256,137 (256,137-256,137) | 0.0 ( 0.00-0.00) | 2,809 ( 2,809- 2,809) |

522

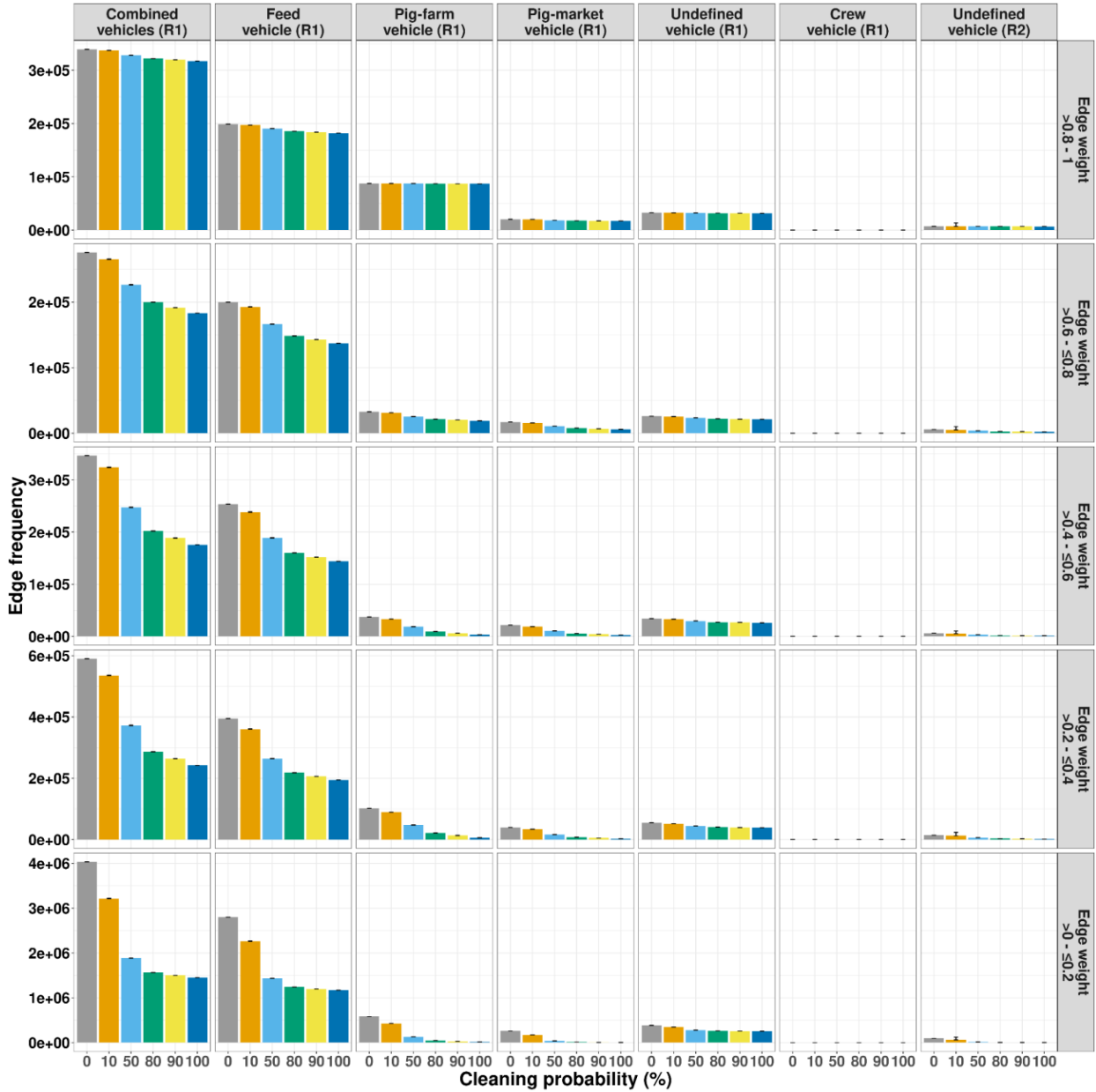

**Figure S16. Distribution of edge weight from ten different reconstructed vehicle contact networks using a VBD of 50 meters and a VVT of 5 minutes and six cleaning probabilities.** Bar graph represent the median values for each clean probability and error line the minimum and maximum ranges for each distribution.

537 The vehicle flow among farm types is available in figures S17 to S30, farm types were grouped
538 accordingly to the classification available in Table S15.
539
540 **Table S15.** Farm unit types grouped.

| Farm unit type | Grouped |
|---|---|
| Sow | Sow |
| Sow-nursery | Sow |
| Sow-nursery-isolation | Sow |
| Gilt-sow | Sow |
| Gilt-sow-nursery | Sow |
| Gilt-sow-nursery-boar stud | Sow |
| Gilt-sow-boar stud | Sow |
| Gilt-sow-nursery-finisher | Sow |
| Nursery | Nursery |
| Gilt-nursery | Nursery |
| Finisher | Finisher |
| Gilt-finisher | Finisher |
| Gilt | GDU |
| Isolation | Isolation |
| Boar stud | Boar stud |
| Wean to finish | Wean-to-finish |
| Wean to finish-finisher | Wean-to-finish |
| Nursery-finisher | Wean-to-finish |
| Nursery-wean to finish-finisher | Wean-to-finish |
| Sow-wean to finish | Farrow-to-finish |
| Sow-finisher | Farrow-to-finish |
| Sow-nursery-finisher | Farrow-to-finish |
| Sow-nursery-finisher-isolation | Farrow-to-finish |
| Gilt-sow-wean to finish | Farrow-to-finish |

| Gilt-sow-wean to finish-boar stud | Farrow-to-finish |
|---|---|
| Gilt-sow-finisher-boar stud | Farrow-to-finish |
| Gilt-sow-nursery-finisher-isolation | Farrow-to-finish |

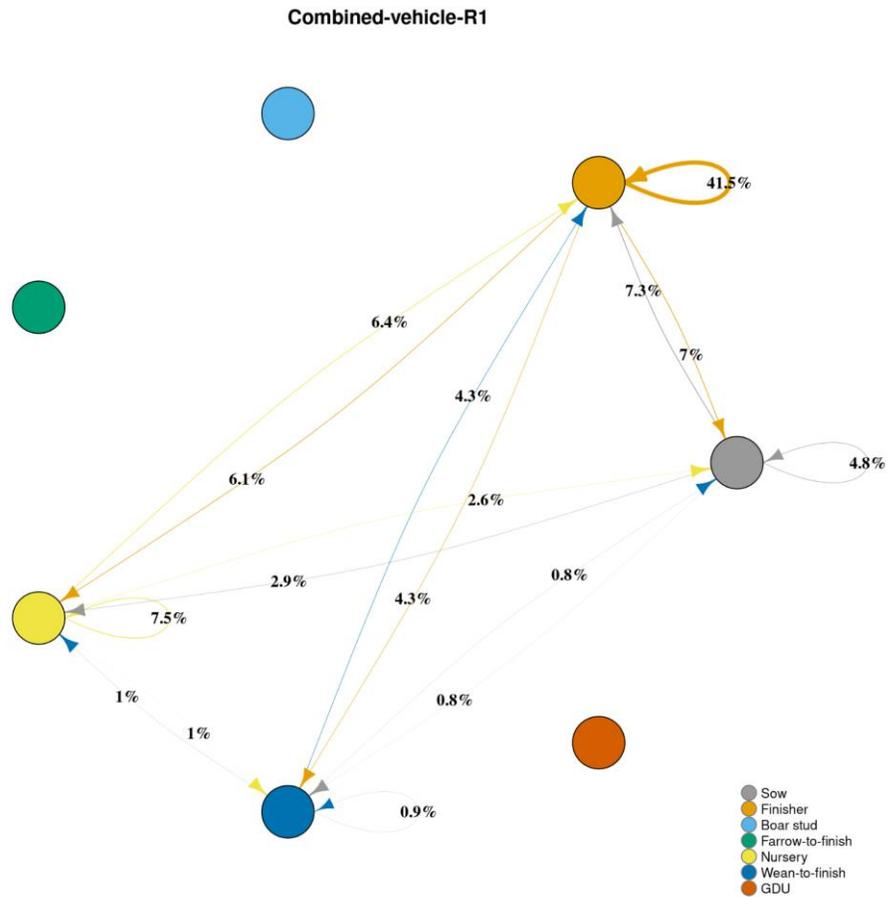

**Figure S17. Flow of vehicle movement among farm types**. Vehicle network reconstructed using a cleaning efficacy of 0%. Edge thickness and values represent the proportion of movement between two farm types; edges with a proportion lower than 0.1% were excluded.

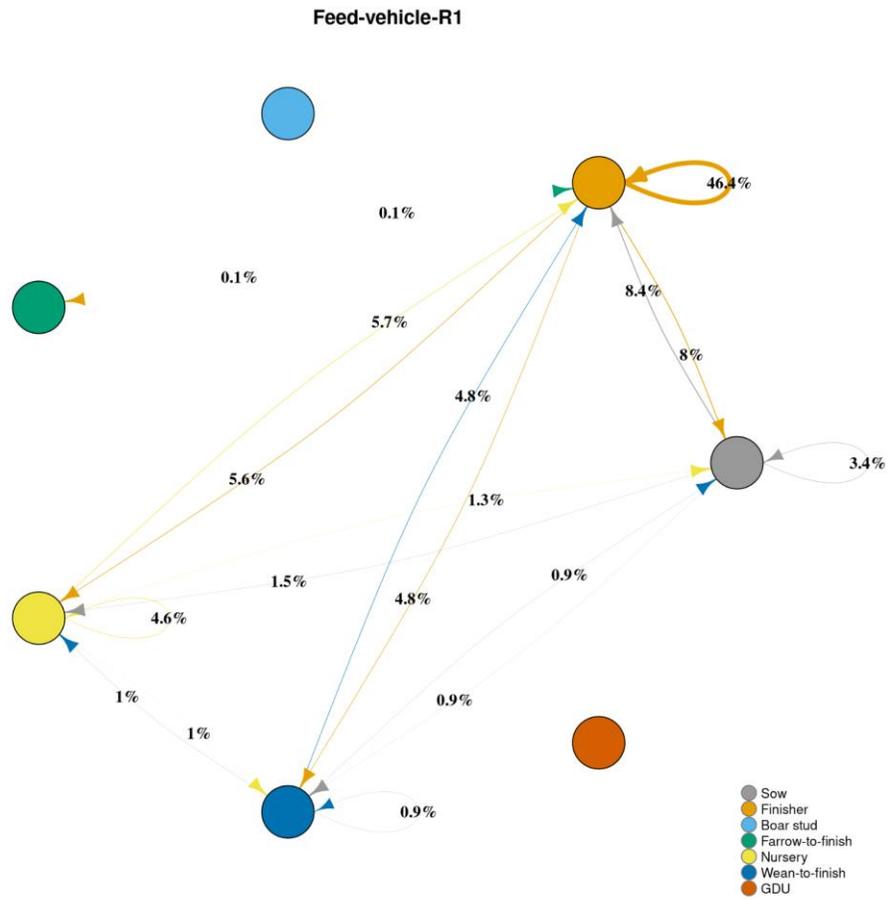

**Figure S18. Flow of vehicle movement among farm types**. Vehicle network reconstructed using a cleaning efficacy of 0%. Edge thickness and values represent the proportion of movement between two farm types; edges with a proportion lower than 0.1% were excluded.

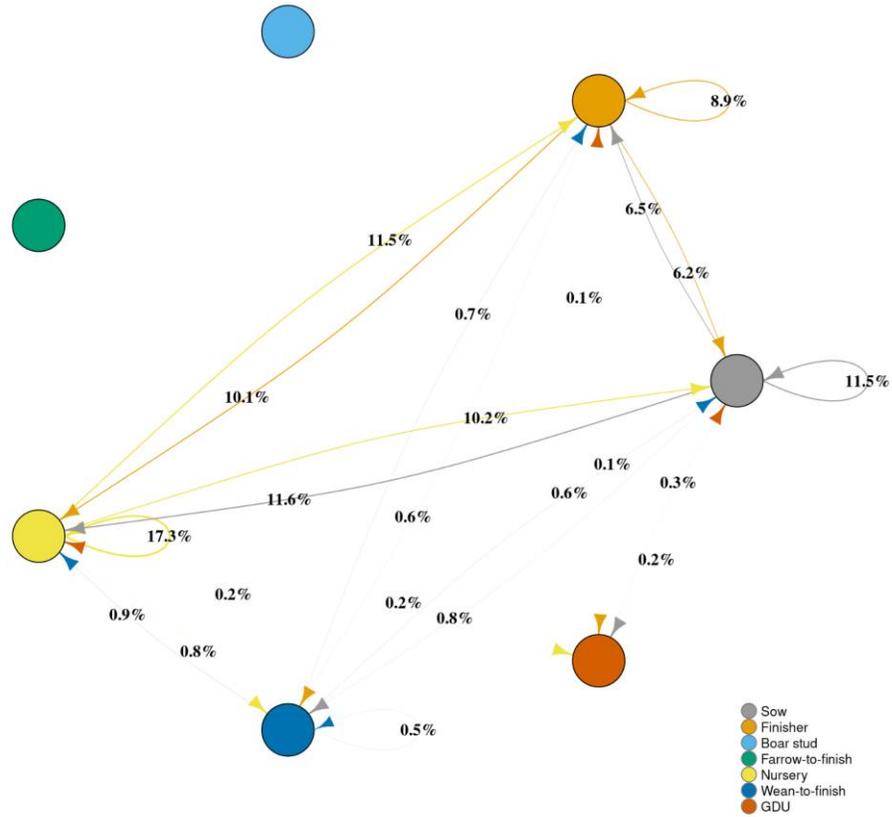

551
552   **Figure S19. Flow of vehicle movement among farm types**. Vehicle network reconstructed using a
553   cleaning efficacy of 0%. Edge thickness and values represent the proportion of movement between two
554   farm types; edges with a proportion lower than 0.1% were excluded.
555
556

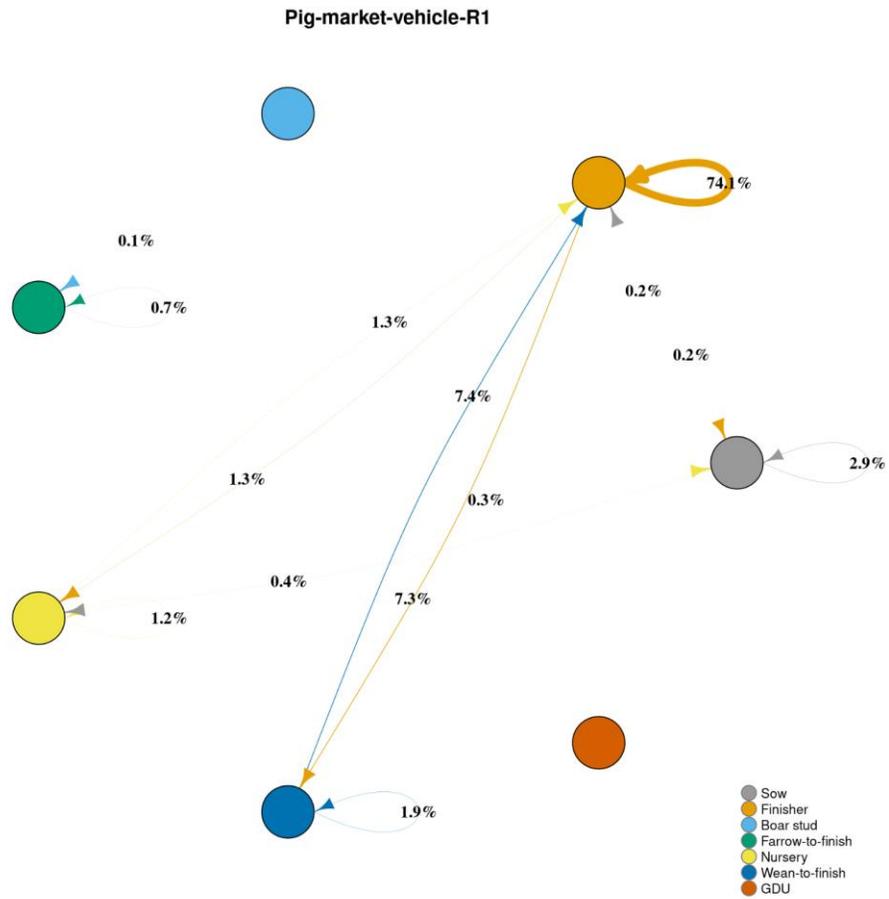

557
558    **Figure S20. Flow of vehicle movement among farm types**. Vehicle network reconstructed using a
559    cleaning efficacy of 0%. Edge thickness and values represent the proportion of movement between two
560    farm types; edges with a proportion lower than 0.1% were excluded.
561

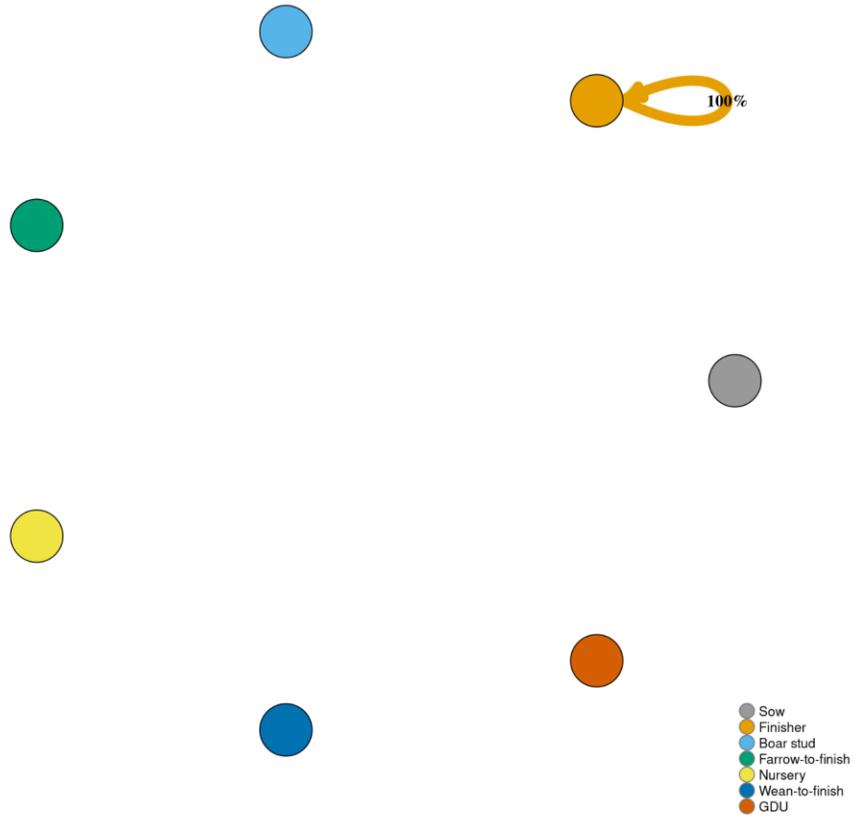

**Figure S21. Flow of vehicle movement among farm types**. Vehicle network reconstructed using a cleaning efficacy of 0%. Edge thickness and values represent the proportion of movement between two farm types; edges with a proportion lower than 0.1% were excluded.

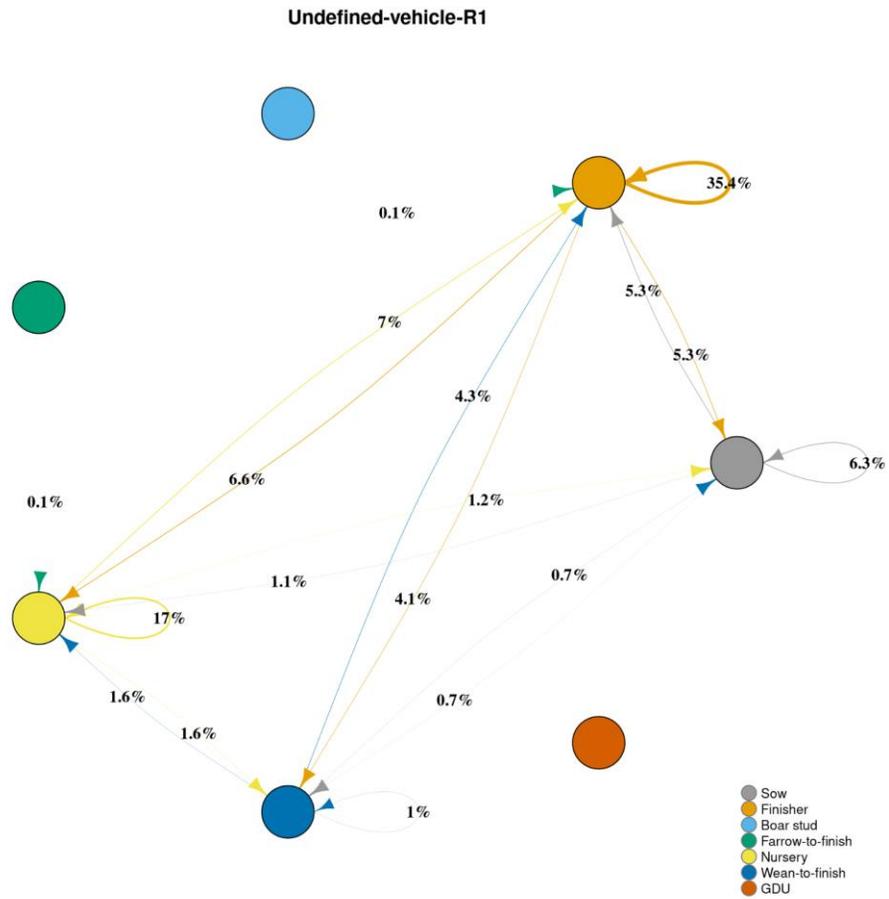

567
568 **Figure S22. Flow of vehicle movement among farm types**. Vehicle network reconstructed using a
569 cleaning efficacy of 0%. Edge thickness and values represent the proportion of movement between two
570 farm types; edges with a proportion lower than 0.1% were excluded.
571

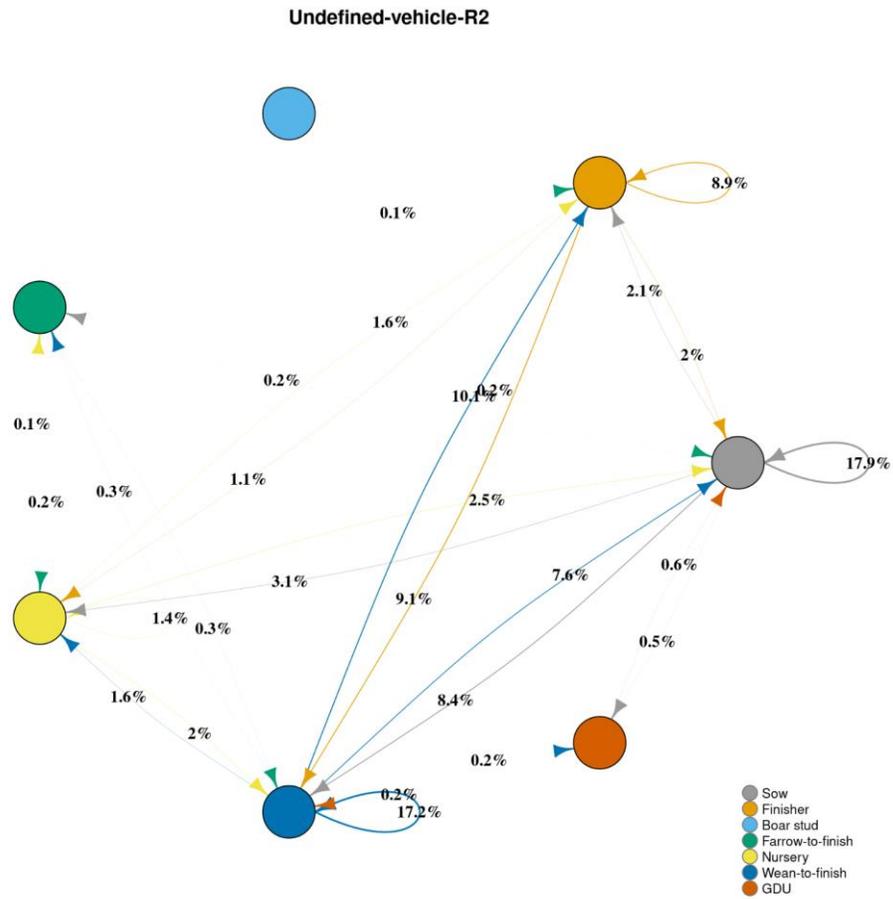

572
573  **Figure S23. Flow of vehicle movement among farm types**. Vehicle network reconstructed using a
574  cleaning efficacy of 0%. Edge thickness and values represent the proportion of movement between two
575  farm types; edges with a proportion lower than 0.1% were excluded.
576

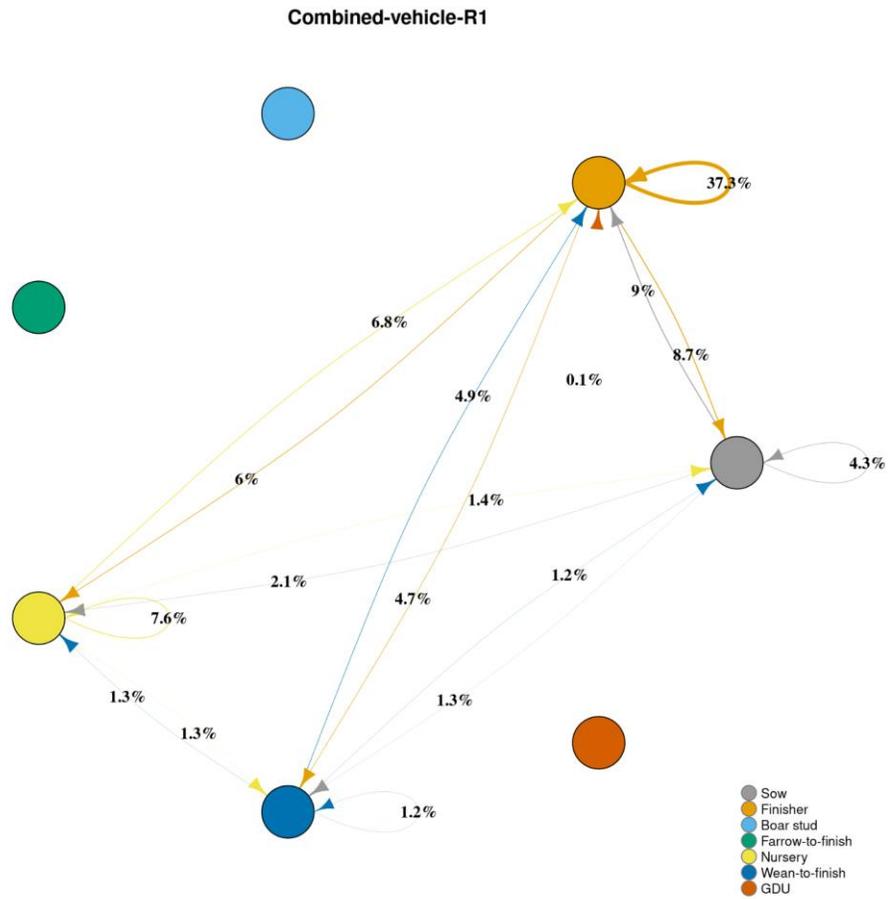

Figure S24. Flow of vehicle movement among farm types. Vehicle network reconstructed using a cleaning efficacy of 100%. Edge thickness and values represent the proportion of movement between two farm types; edges with a proportion lower than 0.1% were excluded.

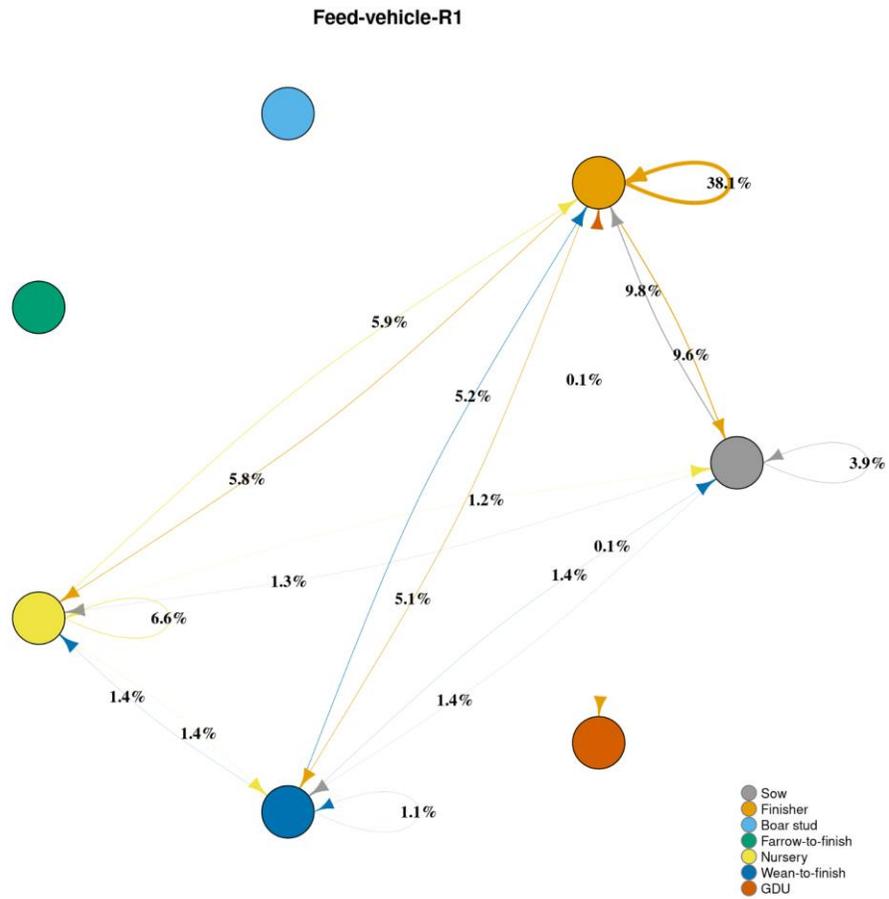

582
583  **Figure S25. Flow of vehicle movement among farm types**. Vehicle network reconstructed using a
584  cleaning efficacy of 100%. Edge thickness and values represent the proportion of movement between two
585  farm types; edges with a proportion lower than 0.1% were excluded.
586

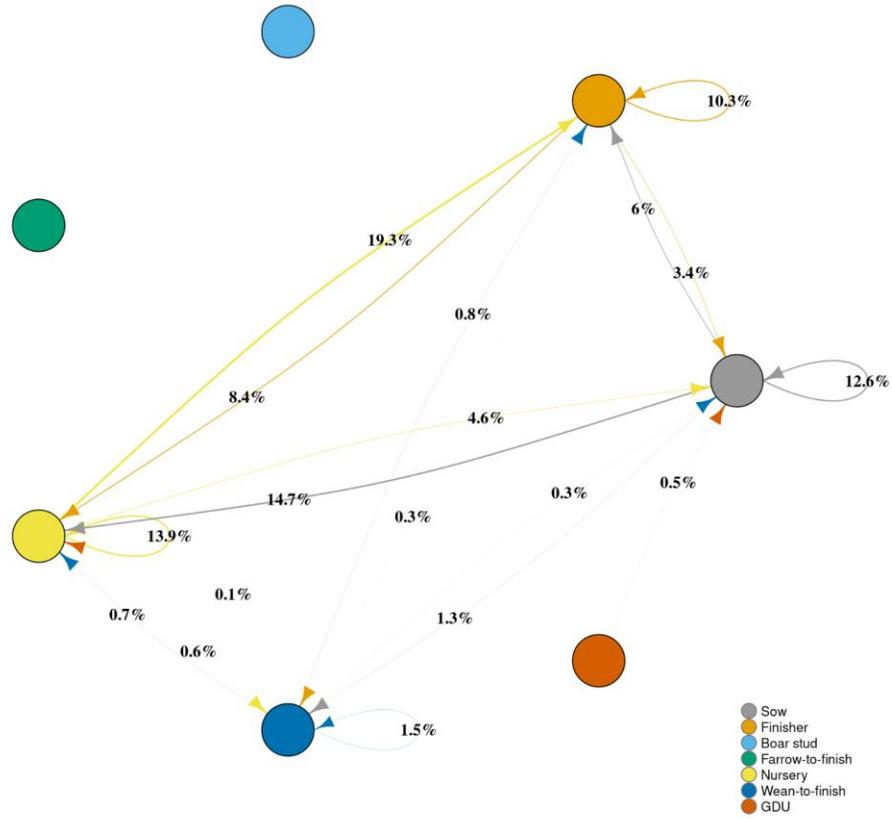

Figure S26. Flow of vehicle movement among farm types. Vehicle network reconstructed using a cleaning efficacy of 100%. Edge thickness and values represent the proportion of movement between two farm types; edges with a proportion lower than 0.1% were excluded.

**Figure S27. Flow of vehicle movement among farm types**. Vehicle network reconstructed using a cleaning efficacy of 100%. Edge thickness and values represent the proportion of movement between two farm types; edges with a proportion lower than 0.1% were excluded.

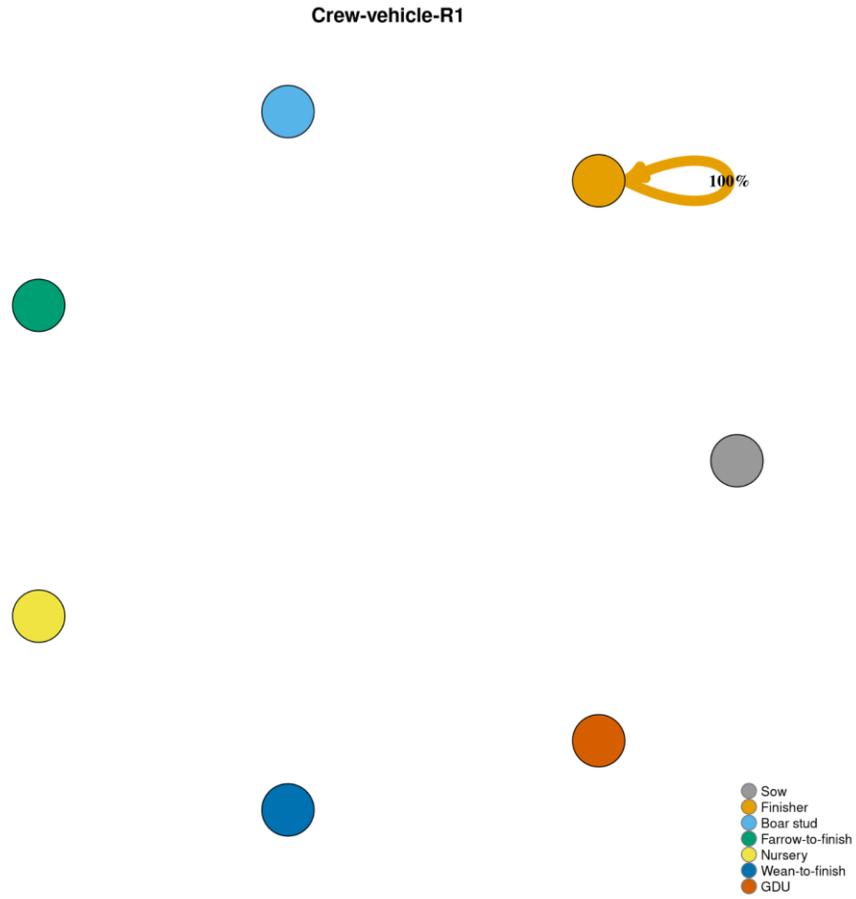

597
598 **Figure S28. Flow of vehicle movement among farm types**. Vehicle network reconstructed using a
599 cleaning efficacy of 100%. Edge thickness and values represent the proportion of movement between two
600 farm types; edges with a proportion lower than 0.1% were excluded.
601

Figure S29. Flow of vehicle movement among farm types. Vehicle network reconstructed using a cleaning efficacy of 100%. Edge thickness and values represent the proportion of movement between two farm types; edges with a proportion lower than 0.1% were excluded.

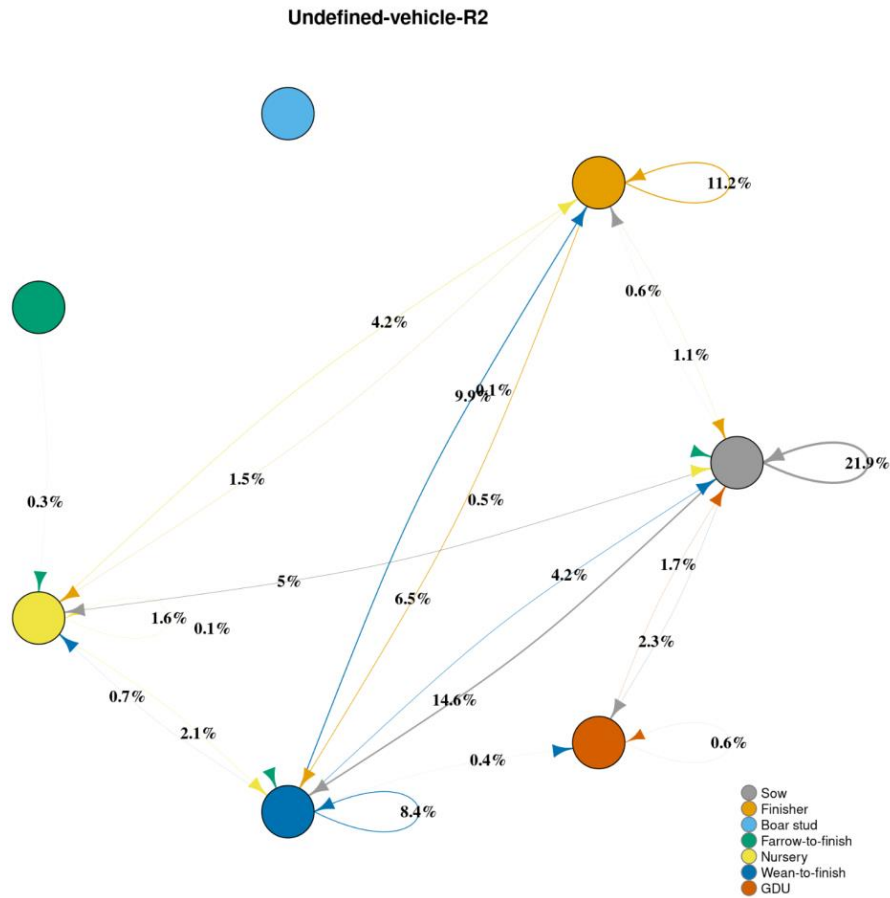

**Figure S30. Flow of vehicle movement among farm types**. Vehicle network reconstructed using a cleaning efficacy of 100%. Edge thickness and values represent the proportion of movement between two farm types; edges with a proportion lower than 0.1% were excluded.